\documentclass[a4paper,DIV=12,11pt]{scrartcl}
\usepackage{mathtools}
\usepackage[pdftex,colorlinks,linkcolor=blue,citecolor=blue,urlcolor=blue]{hyperref}
\usepackage{amssymb}
\usepackage{amsthm}
\usepackage{stix}
\usepackage{inconsolata}
\usepackage{bm}
\usepackage{mleftright}
\mleftright
\usepackage{cite}

\makeatletter
\renewcommand{\bib@heading}{%
  \begin{center}
    \textsc{\refname}
  \end{center}
  \@mkboth{\MakeMarkcase{\refname}}{\MakeMarkcase{\refname}}%
  \small}
\renewcommand{\@biblabel}[1]{\textbf{#1.}}
\makeatother
\let\oldbibliography\thebibliography
\renewcommand{\thebibliography}[1]{%
  \nopagebreak
  \oldbibliography{#1}%
  \setlength{\parskip}{0pt}%
  \setlength{\itemsep}{0pt plus 0.3ex}}

\theoremstyle{remark}
\newtheorem{remark}{Remark}[section]
\setkomafont{section}{\rmfamily\bfseries}
\setkomafont{subsection}{\rmfamily\bfseries}

\def\rme{\mathrm{e}}
\def\rmi{\mathrm{i}}
\def\diag{\mathop{\mathrm{diag}}}
\def\sgn{\mathop{\mathrm{sgn}}}
\def\etal{\textit{et al}}

\begin{document}
\suppressfloats
\begin{center}
  {\bfseries\Large
    Another generalization of the box--ball system with many kinds of balls
  }
  \vskip1em
  \textsc{Kazuki Maeda}\\
  {\itshape
    Department of Mathematical Sciences, School of Science and Technology,\\
    Kwansei Gakuin University, 2-1 Gakuen, Sanda, Hyogo 669-1337, Japan
  }\\
  Email: kmaeda@kmaeda.net
\end{center}

\begin{abstract}
  \noindent
  A cellular automaton that is a generalization of the box--ball system
  with either many kinds of balls or finite carrier capacity is proposed
  and studied through two discrete integrable systems:
  nonautonomous discrete KP lattice
  and nonautonomous discrete two-dimensional Toda lattice.
  Applying reduction technique and ultradiscretization
  procedure to these discrete systems, we derive
  two types of time evolution equations of the proposed
  cellular automaton, and particular solutions to the ultradiscrete equations.
\end{abstract}

\begin{quotation}
\end{quotation}

\section{Introduction}
In this paper, we propose a novel soliton cellular automaton
that is a generalization of the box--ball system (BBS)
with either many kinds of balls or finite carrier capacity,
but not same as the known BBS with both rules
proposed by Hatayama \etal.~\cite{hatayama2001taa}.
It is known that the time evolution equations of the BBSs
are derived by applying a limiting procedure called ultradiscretization
to discrete integrable systems~\cite{tokihiro1996fse}.
There are two types of evolution equations for the BBS.
One is derived from the discrete KdV lattice which corresponds to
the Euler representation and another is from the discrete Toda lattice
with finite boundary condition~\cite{nagai1999sca} which corresponds to
the ``difference form'' of the Lagrange representation, we call
the \emph{finite Toda representation} in this paper.
Note that these Euler--Lagrange notions for cellular automata
come from hydrodynamics~\cite{matsukidaira2003elc}.
In the previous paper~\cite{maeda2017nuh},
it was clarified that the BBS with both many kinds of balls and finite carrier capacity,
originally proposed and analyzed through the $(M+1)$-reduced nonautonomous discrete
KP lattice (nd-KP lattice)~\cite{tokihiro2000bbs},
which corresponds to the Euler representation,
also has the finite Toda representation derived from the nonautonomous
discrete hungry Toda lattice (ndh-Toda lattice).
The proposed cellular automaton also has a simple evolution rule
and comes from the nd-KP lattice and the ndh-Toda lattice via ultradiscretization.
Furthermore, there is an interesting particular solution especially for
the finite Toda representation.

The paper is organized as follows.
In Section~\ref{sec:cellular-automaton},
we explain the time evolution rule of the cellular automaton
studied in this paper, and write two types of time evolution equations on
the min-plus algebra,
i.e. the Euler representation and the finite Toda representation.
In Section~\ref{sec:euler-representation},
we show that the time evolution equation of the Euler representation of
the cellular automaton written in Section~\ref{sec:cellular-automaton}
is derived from the nd-KP lattice by applying
reduction and ultradiscretization procedures.
Imposing the reduction condition to an $N$-soliton solution
of the nd-KP lattice, we also give an $N$-soliton solution
to the Euler representation equation.
In Section~\ref{sec:finite-toda-repr},
we show that the time evolution equation of the finite Toda representation of
the cellular automaton written in Section~\ref{sec:cellular-automaton}
is derive from the ndh-Toda lattice with finite lattice condition
by applying ultradiscretization procedures.
We also give a particular solution to the finite Toda representation.
Section~\ref{sec:concluding-remarks} is devoted to concluding remarks.

\section{A cellular automaton}\label{sec:cellular-automaton}

\subsection{Definition}
Let us consider a finite-state machine that moves in one direction on an infinite tape.
The machine has a state $(V^{(0)}, V^{(1)}, \dots, V^{(M)}) \in \mathbb Z^{M+1}$
and its initial state is $(0, S, S, \dots, S)$, where $S$ is a positive integer or $+\infty$.
We call the state of $V^{(k)}$ \emph{rewritable times of $k$}.
The tape is divided into infinitely many cells and an integer from 0 to $M$ is
written on each cell. Here, we assume that there are a finite number of positive integers
on the tape. The moving machine reads the integer at each cell,
and if the integer is $k$ and $V^{(k)}\ge 1$,
then the machine rewrites the integer $k$ on the cell to $k-1$,
subtracts $1$ from $V^{(k)}$, and adds $1$ to $V^{(k-1)}$,
where $V^{(-1)}=V^{(M)}$ identically; if the integer is $k$ and $V^{(k)}=0$,
then do nothing.

\begin{figure}[t]
  \centering
  \includegraphics[width=\textwidth]{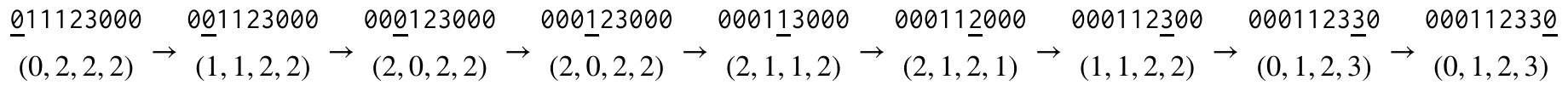}
  \caption{Step-by-step illustration of the rewriting rule
    of the case in which $M=3$ and $S^{(t)}=2$.
    Underline indicates the position of the machine.}
  \label{fig:illust}
\end{figure}

For example, let us consider the case where $M=3$, $S=2$,
and the given tape is
\begin{equation}\label{eq:initial-state-ex-ca}
  \texttt{...000111230001130000...}
\end{equation}
(See also Fig.~\ref{fig:illust}.)
Hereafter, we suppose that the machine on the tape moves from left to right.
Since the initial state of the machine is $(0, 2, 2, 2)$,
i.e. the rewritable times of $0$ is zero,
the machine do nothing until reaching the first `\texttt{1}'.
The machine rewrites the first and second `\texttt{1}'s to `\texttt{0}'s,
and transits its state as $(0, 2, 2, 2)\to (1, 1, 2, 2)\to (2, 0, 2, 2)$.
Now the rewritable times of $1$ becomes zero.
Therefore, the third `\texttt{1}' is not rewritten by the machine.
Next, the machine rewrites `\texttt{2}', `\texttt{3}', `\texttt{0}'
to `\texttt{1}', `\texttt{2}', `\texttt{3}', respectively,
and transits its state as $(2, 0, 2, 2)\to (2, 1, 1, 2)\to (2, 1, 2, 1)\to (1, 1, 2, 2)$.
Finally, the next integers `\texttt{001130}' are rewritten to
`\texttt{300123}' and the state of the machine becomes $(0, 0, 3, 3)$.
After that, the machine does not rewrite remaining infinite `\texttt{0}'s.
We obtain the tape
\begin{equation*}
  \texttt{...000001123300123000...}
\end{equation*}

We introduce a discrete time variable $t$ as the number of iterations of the process above.
We can choose the parameter $S$ at each $t$, write $S^{(t)}$.
\begin{figure}
  \centering
  \includegraphics[width=\textwidth]{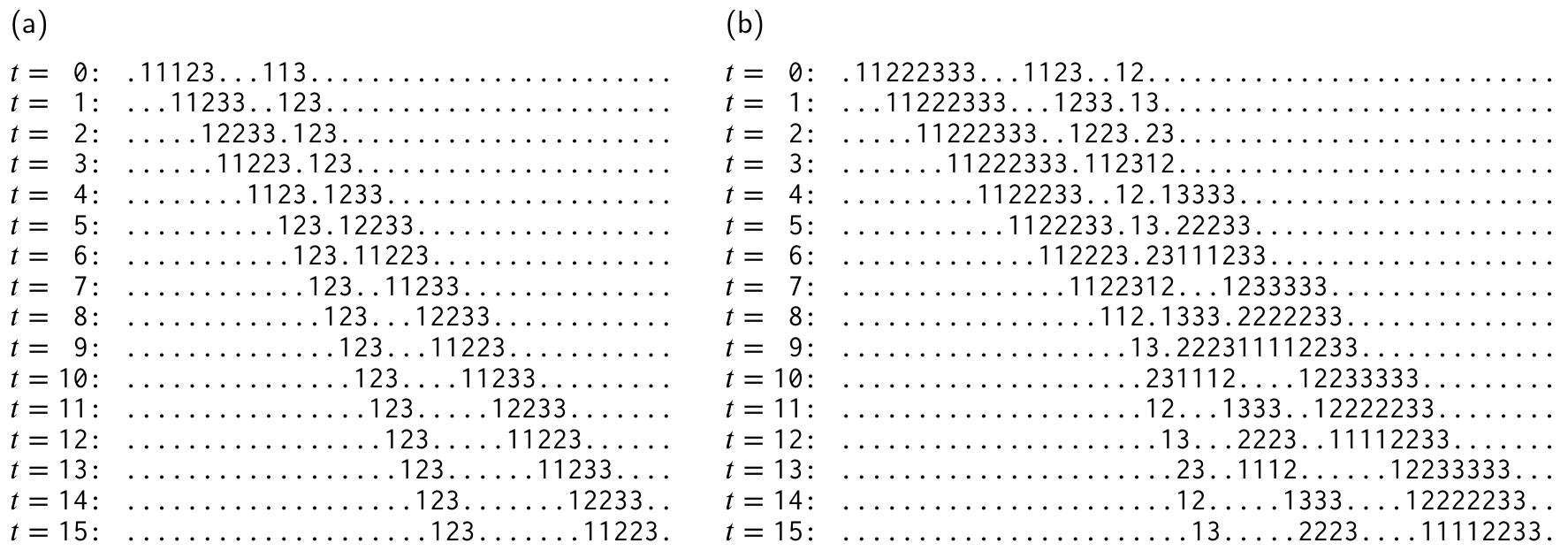}
  \caption{Examples of the time evolution of the proposed cellular automaton
    of the case in which $M=3$ and $S^{(t)}=2$ for all $t$.}
  \label{fig:example-ca}
\end{figure}
Fig.~\ref{fig:example-ca}~(a) shows an example of the time evolution
in which the initial values are given by \eqref{eq:initial-state-ex-ca},
where `\texttt{.}' is used instead of `\texttt{0}'.
A more complicated example is shown in Fig.~\ref{fig:example-ca}~(b).
In these examples, we can observe that the blocks of positive integers
propagate and interact with each other like solitons.
The observation may be reminiscent of the BBS
with many kinds of balls and finite carrier capacity.
In fact, the proposed cellular automaton is a generalization of the BBS.

\subsection{Time evolution equation of the Euler representation}

Let
\begin{equation*}
  U^{(k, t)}_n=
  \begin{cases*}
    1 & if an integer $k$ is written on the $n$th cell at time $t$,\\
    0 & otherwise,
  \end{cases*}
\end{equation*}
and $V^{(k, t)}_n \in \{0, 1, \dots, MS^{(t)}\}$ be
the rewritable times of $k$ just before the machine reads the integer
on the $n$th cell from time $t$ to $t+1$.
Then, we have
\begin{equation*}
  \min(U^{(k, t)}_n, V^{(k, t)}_n)=
  \begin{cases*}
    1 & if $U^{(k, t)}_n=1$ and $V^{(k, t)}_n\ge 1$,\\
    0 & otherwise.
  \end{cases*}
\end{equation*}
Hence, we can write the time evolution equation of
the proposed cellular automaton as
\begin{subequations}\label{eq:evol-eq-ca}
  \begin{align}
    U^{(k, t+1)}_n&=U^{(k, t)}_n-X^{(k, t)}_n+X^{(k+1, t)}_n,\label{eq:evol-eq-ca-u}\\
    V^{(k, t)}_{n+1}&=V^{(k, t)}_n-X^{(k, t)}_n+X^{(k+1, t)}_n,\label{eq:evol-eq-ca-v}\\
    X^{(k, t)}_n&=\min(U^{(k, t)}_n, V^{(k, t)}_n),
  \end{align}
  where $U^{(k+M+1, t)}_n=U^{(k, t)}_n$ and $V^{(k+M+1, t)}_n=V^{(k, t)}_n$ for all
  $k$, $t$ and $n$.
  The boundary condition is given by
  \begin{equation}
    U^{(k, t)}_n=
    \begin{cases*}
      1 & if $k=0$,\\
      0 & if $k=1, 2, \dots, M$,
    \end{cases*}\quad
    V^{(k, t)}_n=
    \begin{cases*}
      0 & if $k=0$,\\
      S^{(t)} & if $k=1, 2, \dots, M$.
    \end{cases*}
  \end{equation}
  for $n \ll -1$.
\end{subequations}
The relations
\begin{equation}\label{eq:constraint}
  \sum_{k=0}^M U^{(k, t)}_n=1,\quad
  \sum_{k=0}^M V^{(k, t)}_n=MS^{(t)}
\end{equation}
always hold obviously.

\begin{remark}\label{rem:m1}
  If $M=1$, then $U^{(0, t)}_n=1-U^{(1, t)}_n$ and
  $V^{(1, t)}_n=S^{(t)}-V^{(0, t)}_n$ hold by \eqref{eq:constraint}.
  Hence, the time evolution equation \eqref{eq:evol-eq-ca-u} is rewritten as
  \begin{align}
    U^{(1, t+1)}_n
    &=U^{(1, t)}_n-\min(U^{(1, t)}_n, S^{(t)}-V^{(0, t)}_n)+\min(1-U^{(1, t)}_n, V^{(0, t)}_n)\nonumber\\
    &=\min(1-U^{(1, t)}_n, V^{(0, t)}_n)+\max(0, U^{(1, t)}_n+V^{(0, t)}_n-S^{(t)}),\label{eq:u-former}
  \end{align}
  where we used the following simple formulae
  \begin{gather*}
    A+\min(B, C)=\min(A+B, A+C),\\
    -\min(-A, -B)=\max(A, B).
  \end{gather*}
  Subtraction of \eqref{eq:evol-eq-ca-u} from \eqref{eq:evol-eq-ca-v} yields
  \begin{align}
    V^{(0, t)}_{n+1}
    &=V^{(0, t)}_n+U^{(1, t)}_n-U^{(1, t+1)}_n\nonumber\\
    &=V^{(0, t)}_{n-1}+U^{(1, t)}_{n-1}-U^{(1, t+1)}_{n-1}+U^{(1, t)}_n-U^{(1, t+1)}_n\nonumber\\
    &=\dots\nonumber\\
    &=\sum_{j=-\infty}^n\left(U^{(1, t)}_j-U^{(1, t+1)}_j\right).\label{eq:v-latter}
  \end{align}
  Substituting \eqref{eq:v-latter} into \eqref{eq:u-former},
  we obtain
  \begin{equation*}
    U^{(1, t+1)}_n=\min\left(1-U^{(1, t)}_n, \sum_{j=-\infty}^{n-1}\left(U^{(1, t)}_j-U^{(1, t+1)}_j\right)\right)+\max\left(0, \sum_{j=-\infty}^{n}U^{(1, t)}_j-\sum_{j=-\infty}^{n-1}U^{(1, t+1)}_j-S^{(t)}\right).
  \end{equation*}
  This equation is well known as the time evolution equation
  of the BBS with carrier capacity $S^{(t)}$.
  In this case, the variable $V^{(0, t)}_n$ denotes the number of balls in the carrier
  at the $n$th box from time $t$ to $t+1$.
\end{remark}

\begin{remark}\label{rem:sinf}
  If $S^{(t)}=+\infty$ for all $t$, then $V^{(k, t)}_n=+\infty$
  and $X^{(k, t)}_n=U^{(k, t)}_n$ hold for $k=1, 2, \dots, M$.
  Then, \eqref{eq:evol-eq-ca-u} is rewritten as
  \begin{align}
    U^{(k, t+1)}_n&=U^{(k+1, t)}_n,\quad k=1, 2, \dots, M-1,\label{eq:evol-eq-ca-u-k}\\
    U^{(M, t+1)}_n&=\min(U^{(0, t)}_n, V^{(0, t)}_n).\nonumber
  \end{align}
  From \eqref{eq:evol-eq-ca-v}, we have
  \begin{align*}
    V^{(0, t)}_{n+1}
    &=V^{(0, t)}_n-\min(U^{(0, t)}_n, V^{(0, t)}_n)+U^{(1, t)}_n\\
    &=V^{(0, t)}_n-U^{(M, t+1)}_n+U^{(1, t)}_n\\
    &=V^{(0, t)}_{n-1}-U^{(M, t+1)}_{n-1}+U^{(1, t)}_{n-1}-U^{(M, t+1)}_n+U^{(1, t)}_n\\
    &=\dots\\
    &=\sum_{j=-\infty}^{n}\left(U^{(1, t)}_{j}-U^{(M, t+1)}_j\right).
  \end{align*}
  Hence, we obtain
  \begin{equation*}
    U^{(M, t+1)}_n=\min\left(1-\sum_{k=1}^M U^{(k, t)}_n, \sum_{j=-\infty}^{n-1}\left(U^{(1, t)}_{j}-U^{(M, t+1)}_j\right)\right).
  \end{equation*}
  Further, from \eqref{eq:evol-eq-ca-u-k},
  \begin{align*}
    U^{(k, t+M)}_n
    &=U^{(k+1, t+M-1)}_n\\
    &=\dots\\
    &=U^{(M, t+k)}_n\\
    &=\min\left(1-\sum_{j=1}^M U^{(j, t+k-1)}_n, \sum_{j=-\infty}^{n-1}\left(U^{(1, t+k-1)}_{j}-U^{(M, t+k)}_j\right)\right)\\
    &=\min\left(1-\sum_{j=1}^{k-1} U^{(j, t+M)}_n-\sum_{j=k}^{M} U^{(j, t)}_n, \sum_{j=-\infty}^{n-1}\left(U^{(k, t)}_{j}-U^{(k, t+M)}_j\right)\right).
  \end{align*}
  This equation is known as the time evolution equation of the BBS
  with $M$ kinds of balls~\cite{tokihiro2000bbs}.
\end{remark}

\subsection{Time evolution equation of the finite Toda representation}

It is known that there is another type of time evolution equation for the BBSs.
Let
\begin{itemize}
\item $Q^{(k, t)}_n$: the number of cells written $k$ in the $n$th block of positive integers at time $t$, $k=1, 2, \dots, M$;
\item $E^{(1, t)}_n$: the number of cells (written $0$) between the $n$th and ($n+1$)st blocks of positive integers at time $t$;
\item $A^{(k, t)}_n$: the rewritable times of $k$ just before the machine arrives at the $n$th block of positive integers from time $t$ to $t+1$, $k=1, 2, \dots, M$;
\item $B^{(1, t)}_n$: the rewritable times of $0$ just before the machine arrives at the $n$th block of $0$, i.e. the block between the $n$th and ($n+1$)st blocks of positive integers, from time $t$ to $t+1$.
\end{itemize}
Then, from time $t$ to $t+1$, $\min(Q^{(k, t)}_n, A^{(k, t)}_n)$ cells written $k$ in the
$n$th block of positive integers are rewritten to $k-1$ by the machine,
where $k=1, 2, \dots, M$, and $\min(B^{(1, t)}_n, E^{(1, t)}_n)$ cells in the
$n$th block of $0$ are rewritten to $M$ by the machine.
Note that the integers in each block of positive integers must be arranged
in ascending order from left to right.
For example, `\texttt{122333112}' is composed of two blocks `\texttt{122333}'
and `112'.
Hence, we can write the time evolution equation of the proposed cellular
automaton as
\begin{subequations}\label{eq:evol-eq-ca-toda}
  \begin{alignat}{2}
    Q^{(k, t+1)}_n&=Q^{(k, t)}_n-\tilde Q^{(k, t)}_n+\tilde Q^{(k+1, t)}_n,&
    A^{(k, t)}_{n+1}&=A^{(k, t)}_n-\tilde Q^{(k, t)}_n+\tilde Q^{(k+1, t)}_n,\\
    E^{(1, t+1)}_n&=E^{(1, t)}_n-\tilde Q^{(M+1, t)}_n+\tilde Q^{(1, t)}_{n+1},&
    B^{(1, t)}_{n+1}&=B^{(1, t)}_n-\tilde Q^{(M+1, t)}_n+\tilde Q^{(1, t)}_{n+1},\\
    \tilde Q^{(k, t)}_n&=\min(Q^{(k, t)}_n, A^{(k, t)}_n),&\quad
    \tilde Q^{(M+1, t)}_n&=\min(B^{(1, t)}_n, E^{(1, t)}_n),
  \end{alignat}
  for $k=1, 2, \dots, M$ and $n=0, 1, 2, \dots, N-1$ with the boundary condition
  \begin{gather}
    A^{(k, t)}_0=S^{(t)},\quad k=1, 2, \dots, M,\\
    B^{(1, t)}_0=\min(Q^{(1, t)}_0, S^{(t)}),\quad
    E^{(1, t)}_{N-1}=+\infty,\quad
  \end{gather}
  for all $t \in \mathbb Z$.
\end{subequations}

\begin{figure}[t]
  \centering
  \includegraphics[width=.8\textwidth]{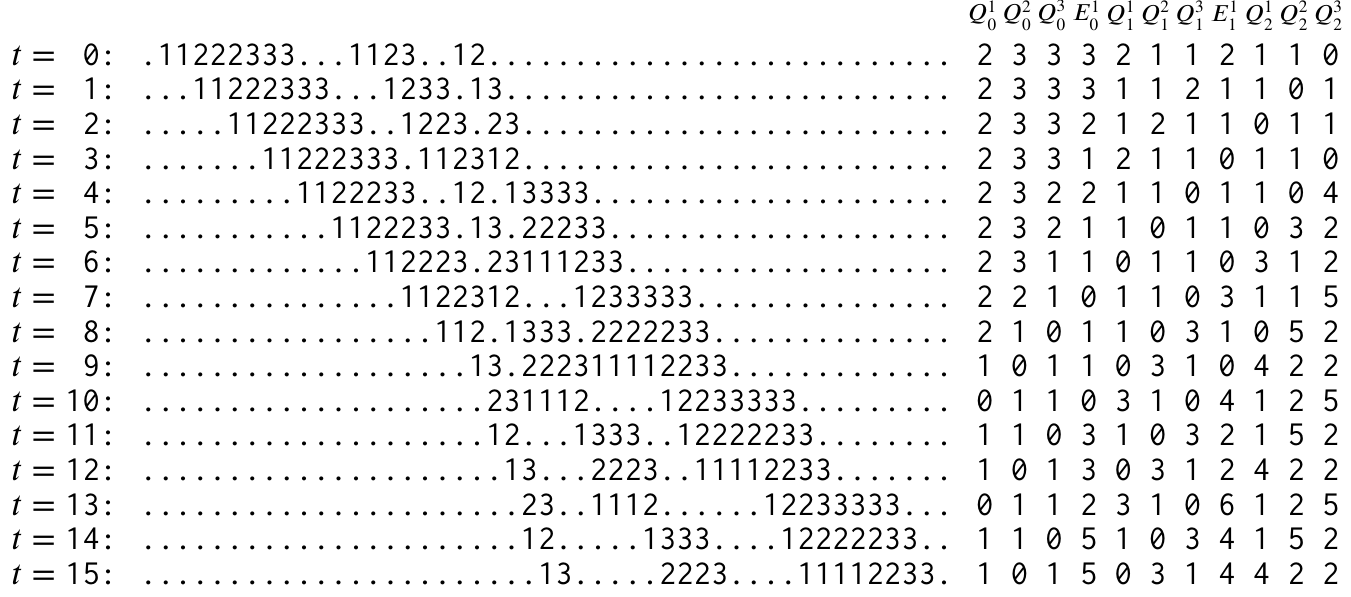}
  \caption{Example of the finite Toda representation of the proposed cellular automaton.
    The left side shows the Euler representation, in which the states and parameters
    ($M=3$ and $S^{(t)}=2$ for all $t$) are same as the ones
    in Fig.~\ref{fig:example-ca}~(b). The right side shows the finite Toda representation
    corresponding to the left side.}
  \label{fig:ex-toda}
\end{figure}

\begin{remark}\label{rem:m1-toda}
  If $M=1$, then we have
  \begin{align*}
    A^{(1, t)}_n
    &=A^{(1, t)}_{n-1}-\tilde Q^{(1, t)}_{n-1}+\tilde Q^{(2 ,t)}_{n-1}\\
    &=A^{(1, t)}_{n-1}-Q^{(1, t)}_{n-1}+Q^{(1, t+1)}_{n-1}\\
    &=A^{(1, t)}_{n-2}-Q^{(1, t)}_{n-2}+Q^{(1, t+1)}_{n-2}-Q^{(1, t)}_{n-1}+Q^{(1, t+1)}_{n-1}\\
    &=\dots\\
    &=A^{(1, t)}_0-\sum_{j=0}^{n-1}\left(Q^{(1, t)}_{j}-Q^{(1, t+1)}_{j}\right)\\
    &=S^{(t)}-\sum_{j=0}^{n-1}\left(Q^{(1, t)}_{j}-Q^{(1, t+1)}_{j}\right).
  \end{align*}
  Hence, we obtain
  \begin{align}
    \tilde Q^{(1, t)}_n
    &=\min\left(Q^{(1, t)}_n, S^{(t)}-\sum_{j=0}^{n-1}\left(Q^{(1, t)}_{j}-Q^{(1, t+1)}_{j}\right)\right)\nonumber\\
    &=\min\left(\sum_{j=0}^n Q^{(1, t)}_j-\sum_{j=0}^{n-1}Q^{(1, t+1)}_j, S^{(t)}\right)-\sum_{j=0}^{n-1}\left(Q^{(1, t)}_{j}-Q^{(1, t+1)}_{j}\right).\label{eq:tq-rel}
  \end{align}
  Similar recursive calculation yields
  \begin{gather*}
    B^{(1, t)}_n=\sum_{j=0}^{n}\tilde Q^{(1, t)}_j-\sum_{j=0}^{n-1}\tilde Q^{(2, t)}_j,\\
    A^{(1, t)}_n
    =S^{(t)}-\sum_{j=0}^{n-1}(\tilde Q^{(1, t)}_j-\tilde Q^{(2, t)}_j)
    =S^{(t)}-B^{(1, t)}_{n}+\tilde Q^{(1, t)}_{n}
    =S^{(t)}-B^{(1, t)}_{n-1}+\tilde Q^{(2, t)}_{n-1}.
  \end{gather*}
  By using these relations, we obtain
  \begin{equation*}
    \sum_{j=0}^{n-1}\left(Q^{(1, t)}_{j}-Q^{(1, t+1)}_{j}\right)
    =B^{(1, t)}_{n}-\tilde Q^{(1, t)}_{n}
    =B^{(1, t)}_{n-1}-\tilde Q^{(2, t)}_{n-1}.
  \end{equation*}
  Substitution of this relation into \eqref{eq:tq-rel} yields
  \begin{gather*}
    Q^{(1, t)}_n-\tilde Q^{(1, t)}_n=\max(0, B^{(1, t)}_{n-1}-\tilde Q^{(2, t)}_{n-1}+Q^{(1, t)}_n-S^{(t)}),\\
    B^{(1, t)}_n=\min(B^{(1, t)}_{n-1}-\tilde Q^{(2, t)}_{n-1}+Q^{(1, t)}_n, S^{(t)}).
  \end{gather*}
  Let $\tilde B^{(1, t)}_n\coloneq Q^{(1, t)}_n-\tilde Q^{(1, t)}_n$.
  Then, the system~\eqref{eq:evol-eq-ca-toda} of the case $M=1$ is rewritten as
  \begin{align*}
    \tilde Q^{(2, t)}_n&=\min(B^{(1, t)}_n, E^{(1, t)}_n),\\
    B^{(1, t)}_{n+1}&=\min(B^{(1, t)}_{n}-\tilde Q^{(2, t)}_{n}+Q^{(1, t)}_{n+1}, S^{(t)}),\\
    \tilde B^{(1, t)}_{n+1}&=\max(0, B^{(1, t)}_{n}-\tilde Q^{(2, t)}_{n}+Q^{(1, t)}_{n+1}-S^{(t)}),\\
    Q^{(1, t+1)}_n&=\tilde Q^{(2, t)}_n+\tilde B^{(1, t)}_n,\\
    E^{(1, t+1)}_n&=E^{(1, t)}_n-\tilde Q^{(2, t)}_n+Q^{(1, t)}_{n+1}-\tilde B^{(1, t)}_{n+1}.
  \end{align*}
  This system is known as the nonautonomous ultradiscrete Toda lattice, which gives
  the time evolution equation of the BBS with finite carrier capacity~\cite{maeda2010bbs}.
\end{remark}

\begin{remark}\label{rem:sinf-toda}
  If $S^{(t)}=+\infty$, then $A^{(k, t)}_n=+\infty$ and $\tilde Q^{(k, t)}_n=Q^{(k, t)}_n$
  hold for all $k=1, 2, \dots, M$, $n=0, 1, \dots, N-1$ and $t \in \mathbb Z$.
  Hence, the system \eqref{eq:evol-eq-ca-toda} becomes
  \begin{align*}
    Q^{(k, t+1)}_n&=Q^{(k+1, t)}_n,\quad k=1, 2, \dots, M-1,\\
    Q^{(M, t+1)}_n&=\tilde Q^{(M+1, t)}_n=\min(B^{(1, t)}_n, E^{(1, t)}_n),\\
    E^{(1, t+1)}_n&=E^{(1, t)}_n-\tilde Q^{(M+1, t)}_n+Q^{(1, t)}_{n+1},\\
    B^{(1, t)}_{n+1}&=B^{(1, t)}_n-\tilde Q^{(M+1, t)}_n+Q^{(1, t)}_{n+1}
  \end{align*}
  with the boundary condition
  \begin{equation*}
    B^{(1, t)}_0=Q^{(1, t)}_0,\quad
    E^{(1, t)}_{N-1}=+\infty.
  \end{equation*}
  Using the recurrence relations recursively,
  we can rewrite $Q^{(M, t+1)}_n$ and $B^{(1, t)}_n$ as
  \begin{gather*}
    Q^{(M, t+1)}_n
    =Q^{(M-1, t+2)}_n
    =Q^{(M-2, t+3)}_n
    =\dots
    =Q^{(1, t+M)}_n,\\
    B^{(1, t)}_n
    =\sum_{j=0}^n Q^{(1, t)}_j-\sum_{j=0}^{n-1} Q^{(1, t+M)}_j,
  \end{gather*}
  respectively. Therefore, we obtain
  \begin{align*}
    Q^{(1, t+M)}_n&=\min\left(\sum_{j=0}^n Q^{(1, t)}_j-\sum_{j=0}^{n-1} Q^{(1, t+M)}_j, E^{(1, t)}_n\right),\\
    E^{(1, t+1)}_n&=E^{(1, t)}_n-Q^{(1, t+M)}_n+Q^{(1, t)}_{n+1}.
  \end{align*}
  This system is known as the ultradiscrete hungry Toda lattice,
  which gives the time evolution equation of the BBS with $M$ kinds
  of balls~\cite{tokihiro1999pos}.
\end{remark}

\begin{remark}
  Remarks~\ref{rem:m1}, \ref{rem:sinf}, \ref{rem:m1-toda} and \ref{rem:sinf-toda}
  say that the proposed cellular automaton is a generalization of
  the BBS with either finite carrier capacity or many kinds of balls.
  However, the proposed cellular automaton is \emph{not}
  the known BBS with \emph{both} finite carrier capacity and many kinds of balls.
  Figs.~\ref{fig:ex-toda-another} and \ref{fig:ex-toda-both} give a comparison
  between the proposed cellular automaton and the BBS with both finite carrier capacity
  and many kinds of balls. In the two examples, the same initial state is given,
  the parameters in Fig.~\ref{fig:ex-toda-another}
  are chosen as $M=3$ and $S^{(t)}=2$ for all $t$,
  and carrier capacity in Fig.~\ref{fig:ex-toda-both} is chosen as $MS^{(t)}=6$ for all $t$.
  We can observe that the states in Fig.~\ref{fig:ex-toda-another} at time $t=3, 6, 9, 12, 15, \dots$
  are very similar, but slightly different,
  to the states in Fig.~\ref{fig:ex-toda-both} at time $t=1, 2, 3, 4, 5, \dots$, respectively.
  On the other hand, we can find that there are initial states and parameter choices
  for which the states after evolution coincide with each other.
  These observations may suggest that there exists a connection between
  the proposed cellular automaton and the BBS with both finite carrier capacity
  and many kinds of balls.

  \begin{figure}[t]
    \centering
    \includegraphics[width=.8\textwidth]{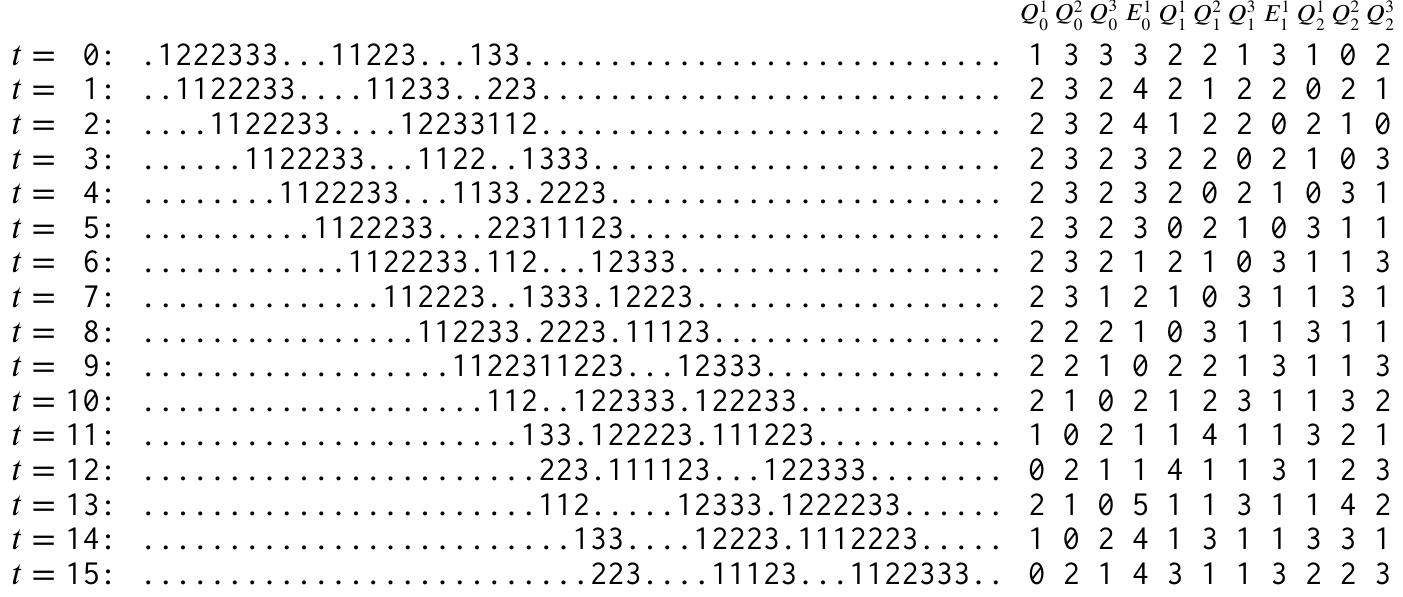}
    \caption{Another example of the correspondence between the Euler representation
    and the finite Toda representation of the proposed cellular automaton.
    The parameters are chosen as $M=3$ and $S^{(t)}=2$ for all $t$.}
    \label{fig:ex-toda-another}

    \vspace{1\baselineskip}

    \includegraphics[width=.8\textwidth]{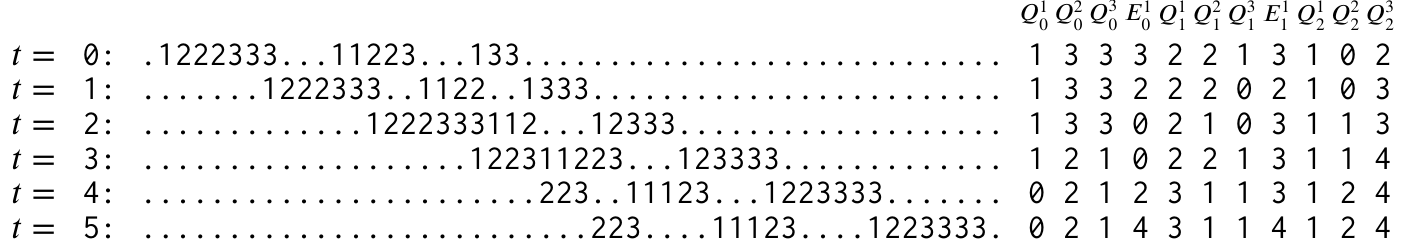}
    \caption{Example of the correspondence between the Euler representation
    and the finite Toda representation of the BBS with both finite carrier capacity
    and many kinds of balls. Carrier capacity is $6$ for all $t$ and
    the number of kinds of balls is $3$.
    The initial state is same as in Fig.~\ref{fig:ex-toda-another}.}
    \label{fig:ex-toda-both}
  \end{figure}

\end{remark}

\section{Euler representation}\label{sec:euler-representation}

\subsection{From the nd-KP lattice}
First, we give a brief exposition on the nd-KP lattice~\cite{willox1997dbd}.
Let us consider the Lax pair
\begin{subequations}\label{eq:ndkp-lax}
  \begin{align}
    \alpha^{(k)} \varphi^{(k+1, t)}_n&=\beta_n\varphi^{(k, t)}_{n+1}+(\alpha^{(k)}-\beta_n)\tilde u^{(k, t)}_n \varphi^{(k, t)}_n,\label{eq:ndkp-lax-u}\\
    \gamma^{(t)} \varphi^{(k, t+1)}_n&=\beta_n\varphi^{(k, t)}_{n+1}+(\gamma^{(t)}-\beta_n)\tilde x^{(k, t)}_n\varphi^{(k, t)}_n,\label{eq:ndkp-lax-x}
  \end{align}
where $k, n, t \in \mathbb Z$ are independent variables,
$\alpha^{(k)}$, $\beta_n$ and $\gamma^{(t)}$ are functions of $k$, $n$ and $t$,
respectively, and $\tilde u^{(k, t)}_n$, $\tilde x^{(k, t)}_n$, and $\varphi^{(k, t)}_n$
are some nonzero functions.
The nd-KP lattice is derived as the compatibility condition for
the Lax pair \eqref{eq:ndkp-lax-u} and \eqref{eq:ndkp-lax-x}.
Subtraction of \eqref{eq:ndkp-lax-u} from \eqref{eq:ndkp-lax-x} yields
\begin{equation}
  \gamma^{(t)}\varphi^{(k, t+1)}_n=\alpha^{(k)}\varphi^{(k+1, t)}_n+(\gamma^{(t)}-\alpha^{(k)})\tilde v^{(k, t)}_n \varphi^{(k, t)}_n,
\end{equation}
\end{subequations}
where $v^{(k, t)}_n$ is a function satisfying
\begin{subequations}
\begin{equation}\label{eq:nd-KP-vux}
  (\gamma^{(t)}-\alpha^{(k)})\tilde v^{(k, t)}_n=(\gamma^{(t)}-\beta_n)\tilde x^{(k, t)}_n-(\alpha^{(k)}-\beta_n)\tilde u^{(k, t)}_n.
\end{equation}
In addition, the following relations are derived as the compatibility conditions for
the linear equations~\eqref{eq:ndkp-lax}:
\begin{gather}
  \tilde u^{(k, t+1)}_n \tilde x^{(k, t)}_n=\tilde u^{(k, t)}_n \tilde x^{(k+1, t)}_n,\label{eq:ndKP-ux}\\
  \tilde v^{(k, t)}_{n+1} \tilde x^{(k, t)}_n=\tilde v^{(k, t)}_n \tilde x^{(k+1, t)}_n.\label{eq:ndKP-vx}
\end{gather}
\end{subequations}
Consider the transformation of dependent variables
\begin{equation}\label{eq:ukv-tau}
  \tilde u^{(k, t)}_n=\frac{f^{(k, t)}_n f^{(k+1, t)}_{n+1}}{f^{(k, t)}_{n+1} f^{(k+1, t)}_n},\quad
  \tilde v^{(k, t)}_n=\frac{f^{(k, t)}_n f^{(k+1, t+1)}_n}{f^{(k, t+1)}_n f^{(k+1, t)}_n},\quad
  \tilde x^{(k, t)}_n=\frac{f^{(k, t)}_n f^{(k, t+1)}_{n+1}}{f^{(k, t)}_{n+1} f^{(k, t+1)}_n}.
\end{equation}
Then, these dependent variables indeed satisfy the relations
\eqref{eq:ndKP-ux} and \eqref{eq:ndKP-vx},
and equation \eqref{eq:nd-KP-vux} is transformed into the bilinear equation
of the nd-KP lattice
\begin{equation*}
  (\alpha^{(k)}-\beta_n)f^{(k, t+1)}_n f^{(k+1, t)}_{n+1}
  +(\beta_n-\gamma^{(t)})f^{(k+1, t)}_{n}f^{(k, t+1)}_{n+1}
  +(\gamma^{(t)}-\alpha^{(k)})f^{(k, t)}_{n+1}f^{(k+1, t+1)}_{n}=0.
\end{equation*}
It is known that an $N$-soliton solution of the nd-KP lattice is given by
\begin{gather*}
  f^{(k, t)}_n=1+\sum_{\substack{J\subset\{0, 1, \dots, N-1\}\\J\ne \emptyset}}\left(\prod_{\substack{r_0, r_1 \in J\\r_0<r_1}}\omega_{r_0, r_1}\prod_{r\in J}\tilde\varphi_{r, n}^{(k, t)}\right),\\
  \omega_{r_0, r_1}\coloneq \frac{(p_{r_0}-p_{r_1})(\tilde p_{r_1}-\tilde p_{r_0})}{(p_{r_0}-\tilde p_{r_1})(p_{r_1}-\tilde p_{r_0})},\quad
  \tilde\varphi_{r, n}^{(k, t)}\coloneq \theta_r\prod_{k'=0}^{k-1}\frac{\alpha^{(k')}-\tilde p_r}{\alpha^{(k')}-p_r}\prod_{n'=0}^{n-1}\frac{\beta_{n'}-\tilde p_r}{\beta_{n'}-p_r}\prod_{t'=0}^{t-1}\frac{\gamma^{(t')}-\tilde p_r}{\gamma^{(t')}-p_r},
\end{gather*}
where $p_r$, $\tilde p_r$ and $\theta_r$ are some constants satisfying
$p_{r_0} \ne \tilde p_{r_1}$ for all $r_0, r_1\in \{ 0, 1, \dots, N-1\}$.

Next, we impose the $(M+1)$-reduction condition
\begin{equation}\label{eq:M1-reduc-cond}
  f^{(k+M+1, t)}_n=f^{(k, t)}_n,\quad
  \alpha^{(k+M+1)}=\alpha^{(k)}
\end{equation}
for all $k, n, t$, which is the same condition as the one for deriving
the time evolution equation of the BBS with $M$ kinds of balls.
We define new functions $\delta_n=-1+\beta_n$ and $s^{(t)}=-\gamma^{(t)}$, and set
\begin{equation*}
  \alpha^{(k)}=
  \begin{cases*}
    0 & if $k=0, 1, \dots, M-1$,\\
    1 & if $k=M$.
  \end{cases*}
\end{equation*}
Further, let us define the new dependent variables
\begin{subequations}\label{eq:dep-var}
\begin{gather}
  u^{(k, t)}_n=
  \begin{cases*}
    \delta_n \tilde u^{(M, t)}_n & if $k=0$,\\
    (1+\delta_n)\tilde u^{(k-1, t)}_n & if $k=1, 2, \dots, M$,
  \end{cases*}\
  v^{(k, t)}_n=
  \begin{cases*}
    (1+s^{(t)})\tilde v^{(M, t)}_n & if $k=0$,\\
    s^{(t)}\tilde v^{(k-1, t)}_n & if $k=1, 2, \dots, M$,
  \end{cases*}\\
  x^{(k, t)}_n=(1+\delta_n+s^{(t)})\tilde x^{(k-1, t)}_n.
\end{gather}
\end{subequations}
Then, we obtain the equations
\begin{subequations}\label{eq:evol-iud-eq-ca}
\begin{equation}
  u^{(k, t+1)}_n=u^{(k, t)}_n\frac{x^{(k+1, t)}_n}{x^{(k, t)}_n},\quad
  v^{(k, t)}_{n+1}=v^{(k, t)}_n\frac{x^{(k+1, t)}_n}{x^{(k, t)}_n},\quad
  x^{(k, t)}_n=u^{(k, t)}_n+v^{(k, t)}_n
\end{equation}
with the boundary conditions
\begin{equation}
  u^{(k, t)}_n=
  \begin{cases*}
    \delta_n & if $k=0$,\\
    1+\delta_n & if $k=1, 2, \dots, M$,
  \end{cases*}\quad
  v^{(k, t)}_n=
  \begin{cases*}
    1+s^{(t)} & if $k=0$,\\
    s^{(t)} & if $k=1, 2, \dots, M$,
  \end{cases*}
\end{equation}
\end{subequations}
for $n\ll -1$.
We also have
\begin{equation}\label{eq:constraint-d}
  \prod_{k=0}^{M}u^{(k, t)}_n=\delta_n(1+\delta_n)^M,\quad
  \prod_{k=0}^{M}v^{(k, t)}_n=(s^{(t)})^M(1+s^{(t)})
\end{equation}
for all $n$ and $t$.

\subsection{Particular solutions}
According to the previous studies~\cite{tokihiro2000bbs,hatayama2001taa},
we construct an $N$-soliton solution to \eqref{eq:evol-eq-ca}.
The construction is complicated, but similar to the previous studies;
the only difference is the choice of the parameter $\gamma^{(t)}$.
Therefore, we give only the result here.
Set $u^{(k, t)}_n=\rme^{-U^{(k, t)}_n/\epsilon}$,
$v^{(k, t)}_n=\rme^{-V^{(k, t)}_n/\epsilon}$,
$x^{(k, t)}_n=\rme^{-X^{(k, t)}_n/\epsilon}$,
$\delta_n=\rme^{-\Delta_n/\epsilon}$,
$s^{(t)}=\rme^{-S^{(t)}/\epsilon}$,
and suppose that $\Delta_n\ge 0$ and $S^{(t)}\ge 0$.
Taking the limit $\epsilon \to +0$, then the system~\eqref{eq:evol-iud-eq-ca} goes to
the time evolution equation of Euler representation~\eqref{eq:evol-eq-ca}
and the relations~\eqref{eq:constraint-d} also go to
\eqref{eq:constraint} through the ultradiscretization formula
\begin{equation*}
  \lim_{\epsilon \to +0}-\epsilon\log(\rme^{-A/\epsilon}+\rme^{-B/\epsilon})=\min(A, B).
\end{equation*}

\begin{remark}
  To obtain the time evolution equation~\eqref{eq:evol-eq-ca},
  we should set $\Delta_n=1$ identically.
  In general, the parameter $\Delta_n$ gives the capacity of $n$th box (cell).
  This generalization is easily obtained by changing the initial and boundary conditions
  to
  \begin{equation*}
    \sum_{k=0}^{M} U^{(k, 0)}_n=\Delta_n
  \end{equation*}
  for all $n\in\mathbb Z$ and
  \begin{equation*}
    U^{(k, t)}_n=
    \begin{dcases*}
      \Delta_n & if $k=0$,\\
      0 & if $k=1, 2, \dots, M$,
    \end{dcases*}
  \end{equation*}
  for $n\ll -1$, respectively.
\end{remark}

Soliton solutions are given as follows.
From \eqref{eq:ukv-tau} and \eqref{eq:dep-var},
$f^{(k, t)}_n=\rme^{-F^{(k, t)}_n/\epsilon}$ and $\epsilon\to+0$ yield
\begin{align*}
  U^{(k, t)}_n&=
  \begin{dcases*}
    F^{(M, t)}_n-F^{(M, t)}_{n+1}+F^{(0, t)}_{n+1}-F^{(0, t)}_n+\Delta_n & if $k=0$,\\
    F^{(k-1, t)}_n-F^{(k-1, t)}_{n+1}+F^{(k, t)}_{n+1}-F^{(k, t)}_n & if $k=1, 2, \dots, M$,
  \end{dcases*}\\
  V^{(k, t)}_n&=
  \begin{dcases*}
    F^{(M, t)}_n-F^{(M, t+1)}_n+F^{(0, t+1)}_{n}-F^{(0, t)}_n & if $k=0$,\\
    F^{(k-1, t)}_n-F^{(k-1, t+1)}_n+F^{(k, t+1)}_{n}-F^{(k, t)}_n+S^{(t)} & if $k=1, 2, \dots, M$,
  \end{dcases*}\\
  X^{(k, t)}_n&=F^{(k-1, t)}_n-F^{(k-1, t)}_{n+1}+F^{(k-1, t+1)}_{n+1}-F^{(k-1, t+1)}_n.
\end{align*}
Here, a one-soliton solution to the ($M+1$)-reduced nd-KP lattice is given by
\begin{gather}
  f^{(k, t)}_n=1+\sum_{j=0}^{M-1}\tilde\psi^{(k, t)}_{0, j, n},\label{eq:one-soliton}\\
  \begin{split}
    \tilde\psi^{(k, t)}_{r, j, n}
    &\coloneq \theta_{r, j}\left(\frac{\tilde z_{r, j}}{1-z_r}\right)^{k'}\prod_{n'=0}^{n-1}\frac{1-\tilde z_{r, j}+\delta_{n'}}{z_r+\delta_{n'}}\prod_{t'=0}^{t-1}\frac{\tilde z_{r, j}+s^{(t')}}{1-z_r+s^{(t')}}\nonumber\\
    &=\theta_{r, j}\sum_{l=0}^t \left(\left(\frac{\tilde z_{r, j}}{1-z_r}\right)^{k'+l}(1-z_r)^l \tilde s^{(t)}_{t-l}\right) \prod_{n'=0}^{n-1}\frac{1-\tilde z_{r, j}+\delta_{n'}}{z_r+\delta_{n'}}\prod_{t'=0}^{t-1}\frac{1}{1-z_r+s^{(t')}}\nonumber\\
    &=
    \begin{multlined}[t]
      \theta_{r, j}\sum_{l=0}^t \left(\left(\frac{\tilde z_{r, j}}{1-z_r}\right)^{(k'+l)\bmod M} \left(\frac{z_r}{1-\tilde z_{r, j}}\right)^{\lfloor (k'+l)/M\rfloor} (1-z_r)^l \tilde s^{(t)}_{t-l}\right)\\
      \times\prod_{n'=0}^{n-1}\frac{1-\tilde z_{r, j}+\delta_{n'}}{z_r+\delta_{n'}}\prod_{t'=0}^{t-1}\frac{1}{1-z_r+s^{(t')}},
    \end{multlined}
  \end{split}\nonumber
\end{gather}
where $0<z_r<1$,
$k'=k\bmod (M+1)$,
$\lfloor x\rfloor\coloneq\max\{\, m\in\mathbb Z: m \le x\,\}$ is the floor function,
$\tilde z_{r, 0}, \tilde z_{r, 1}, \dots, \tilde z_{r, M-1} \in \mathbb C$
are the roots of the following algebraic equation except $1-z_r$:
\begin{equation*}
  \left(\frac{z}{1-z_r}\right)^M=\frac{z_r}{1-z},
\end{equation*}
which corresponds to the $(M+1)$-reduction condition~\eqref{eq:M1-reduc-cond},
and
\begin{equation*}
  \tilde s_i^{(t)}\coloneq
  \begin{dcases*}
    1& if $i=0$,\\
    \sum_{0\le j_0<j_1<\dots<j_{i-1}\le t-1}\prod_{l=0}^{i-1}s^{(j_l)}& if $i=1, 2, \dots, t$,\\
    0& otherwise.
  \end{dcases*}
\end{equation*}
Applying the method of the previous studies to \eqref{eq:one-soliton},
we obtain a one-soliton solution to the ultradiscrete system~\eqref{eq:evol-eq-ca}:
\begin{gather*}
  F^{(k, t)}_n=\min\left(0, \tilde\Psi^{(k, t)}_{0, n}(0)\right),\\
  \tilde\Psi^{(k, t)}_{r, n}(i)\coloneq\Theta_r+\min_{l=0, 1, \dots, t}\left(\sum_{j=1}^{(k'+l+i)\bmod M} \zeta_{r, j}+\left(\left\lfloor \frac{k'+l+i}{M}\right\rfloor+i\right) Z_r+\tilde S^{(t)}_{t-l}\right)-\sum_{n'=0}^{n-1}\min(Z_r, \Delta_{n'}),
\end{gather*}
where $Z_r\ge 0$ and $\Theta_r$ are the parameters determine
the size and phase of the soliton, respectively,
$\zeta_{r, 1}, \zeta_{r, 2}, \dots, \zeta_{r, M}$
are nonnegative parameters satisfying $\sum_{j=1}^{M} \zeta_{r, j}=Z_r$, and
\begin{equation*}
  \tilde S^{(t)}_i\coloneq
  \begin{dcases*}
    0 & if $i=0$,\\
    \min_{0\le j_0<j_1<\dots<j_{i-1}\le t-1} \left(\sum_{l=0}^{i-1} S^{(j_l)}\right) & if $i=1, 2, \dots, t$,\\
    +\infty & otherwise.
  \end{dcases*}
\end{equation*}

In general, an $N$-soliton solution to the $(M+1)$-reduced nd-KP lattice is given by
\begin{gather*}
  f^{(k, t)}_n=1+\sum_{\substack{0\le r_0<r_1<\dots<r_{m-1}\le N-1\\ m=1, 2, \dots, N}}\left(\sum_{j_0, j_1, \dots, j_{m-1}=0}^{M-1} \prod_{0\le i_0<i_1\le m-1}\tilde \omega_{r_{i_0}, j_{i_0}; r_{i_1}, j_{i_1}}\prod_{i=0}^m\tilde \psi^{(k, t)}_{r_i, j_i, n}\right),\\
  \tilde\omega_{r_0, j_0; r_1, j_1}
  \coloneq\frac{(z_{r_1}-z_{r_0})(\tilde z_{r_1, j_1}-\tilde z_{r_0, j_0})}{(1-z_{r_0}-\tilde z_{r_1, j_1})(1-z_{r_1}-\tilde z_{r_0, j_0})}.
\end{gather*}
Applying the ultradiscretization procedure to this solution,
we obtain an $N$-soliton solution to the ultradiscrete system~\eqref{eq:evol-eq-ca}:
\begin{equation*}
  F^{(k, t)}_n=\min\left(0, \min_{\substack{0\le r_0<r_1<\dots<r_{m-1}\le N-1\\ m=1, 2, \dots, N}}\left(\sum_{i=0}^{m-1} \tilde\Psi^{(k, t)}_{r_i, n}(i)\right)\right).
\end{equation*}
Note that the parameters must satisfy $Z_0\ge Z_1\ge \dots \ge Z_{N-1}\ge 0$ and
$\zeta_{0, j}\ge \zeta_{1, j}\ge \dots \ge \zeta_{N-1, j}\ge 0$ for $j=1, 2, \dots, M$.
For example, the solution corresponding to Fig.~\ref{fig:ex-toda} is given by setting
$\Theta_0=1$, $\Theta_1=7$, $\Theta_2=13$,
$Z_0=8$, $Z_1=4$, $Z_2=2$,
$\zeta_{0, 1}=2$, $\zeta_{0, 2}=3$, $\zeta_{0, 3}=3$,
$\zeta_{1, 1}=1$, $\zeta_{1, 2}=2$, $\zeta_{1, 3}=1$,
$\zeta_{2, 1}=1$, $\zeta_{2, 2}=0$, $\zeta_{2, 3}=1$,
$\Delta_n=1$ for all $n$, and $S^{(t)}=2$ for all $t$.

\section{Finite Toda representation}\label{sec:finite-toda-repr}

\subsection{$(M, 1)$-biorthogonal polynomials and the ndh-Toda lattice}
First, we prepare the notion of $(M, 1)$-biorthogonal polynomials~\cite{maeda2017nuh}.
We introduce spectral transformations of the $(M, 1)$-biorthogonal polynomials,
and derive the ndh-Toda lattice as their compatibility condition.

Let $\mathcal L^{(k, t)}\colon \mathbb C[z]\to \mathbb C$ be a linear functional,
where $k, t \in \mathbb Z$ are discrete time variables.
Let us consider polynomial sequences $\{\phi^{(k, t)}_n(z)\}_{n=0}^\infty$
and $\{\psi^{(k, t)}_n\}_{n=0}^\infty$ satisfying the following properties:
\begin{enumerate}
\item $\deg\phi^{(k, t)}_n(z)=\deg\psi^{(k, t)}_n(z)=n$;
\item The polynomials $\phi^{(k, t)}_n(z)$ and $\psi^{(k, t)}_n(z)$ are monic;
\item There exists a positive integer $M$ such that
  the \emph{$(M, 1)$-biorthogonal relation} with respect to $\mathcal L^{(k, t)}$
  \begin{equation}\label{eq:m1-biorth-orig}
    \mathcal L^{(k, t)}[\phi^{(k, t)}_m(z)\psi^{(k, t)}_n(z^M)]=h^{(k, t)}_n\delta_{m, n},\quad
    h^{(k, t)}_n\ne 0,\quad
    m, n=0, 1, 2, \dots,
  \end{equation}
  holds, where $\delta_{m, n}$ is the Kronecker delta.
\end{enumerate}

In this paper, we call the polynomial sequences $\{\phi^{(k, t)}_n(z)\}_{n=0}^\infty$
and $\{\psi^{(k, t)}_n(z)\}_{n=0}^\infty$ the pair of monic
\emph{$(M, 1)$-biorthogonal polynomial sequences} with respect to $\mathcal L^{(k, t)}$.
Note that the $(M, 1)$-biorthogonal relation~\eqref{eq:m1-biorth-orig} is equivalent
to
\begin{equation}\label{eq:m1-biorth}
  \mathcal L^{(k, t)}[z^{Mm}\phi^{(k, t)}_n(z)]
  =\mathcal L^{(k, t)}[z^m\psi^{(k, t)}_n(z^M)]
  =h^{(k, t)}_n\delta_{m, n},\quad
  n=0, 1, 2, \dots,\quad
  m=0, 1, \dots, n.
\end{equation}

Let us introduce time evolution into the linear functional by
\begin{equation}\label{eq:time-evol-lf}
  \mathcal L^{(k+1, t)}[z^m]\coloneq \mathcal L^{(k, t)}[z^{m+1}],\quad
  \mathcal L^{(k, t+1)}[z^m]\coloneq \mathcal L^{(k, t)}[z^{m}(z+s^{(t)})],\quad
  m=0, 1, 2, \dots,
\end{equation}
where $s^{(t)}$ is a parameter depending on $t$.
Define the moment of $\mathcal L^{(0, t)}$ by
\begin{equation*}
  \mu^{(t)}_m\coloneq \mathcal L^{(0, t)}[z^m],\quad m=0, 1, 2, \dots.
\end{equation*}
Then we have
\begin{equation*}
  \mathcal L^{(k, t)}[z^m]=\mu^{(t)}_{k+m},\quad
  \mu^{(t+1)}_m=\mu^{(t)}_{m+1}+s^{(t)}\mu^{(t)}_m.
\end{equation*}

By using the $(M, 1)$-biorthogonal relation~\eqref{eq:m1-biorth}
and the moments, one can easily show that the $(M, 1)$-biorthogonal
polynomials have the following determinant representation:
\begin{align*}
  \phi^{(k, t)}_0(z)&=1,&
  \phi^{(k, t)}_n(z)&=\frac{1}{\tau^{(k, t)}_n}
  \begin{vmatrix}
    \mu^{(t)}_k & \mu^{(t)}_{k+M} & \dots & \mu^{(t)}_{k+M(n-1)} & 1\\
    \mu^{(t)}_{k+1} & \mu^{(t)}_{k+1+M} & \dots & \mu^{(t)}_{k+1+M(n-1)} & z\\
    \vdots & \vdots & & \vdots & \vdots\\
    \mu^{(t)}_{k+n-1} & \mu^{(t)}_{k+n-1+M} & \dots & \mu^{(t)}_{k+n-1+M(n-1)} & z^{n-1}\\
    \mu^{(t)}_{k+n} & \mu^{(t)}_{k+n+M} & \dots & \mu^{(t)}_{k+n+M(n-1)} & z^{n}
  \end{vmatrix}
  ,&
  n=1, 2, 3, \dots,\\
  \psi^{(k, t)}_0(z)&=1,&
  \psi^{(k, t)}_n(z)&=\frac{1}{\tau^{(k, t)}_n}
  \begin{vmatrix}
    \mu^{(t)}_k & \mu^{(t)}_{k+M} & \dots & \mu^{(t)}_{k+M(n-1)} & \mu^{(t)}_{k+Mn}\\
    \mu^{(t)}_{k+1} & \mu^{(t)}_{k+1+M} & \dots & \mu^{(t)}_{k+1+M(n-1)} & \mu^{(t)}_{k+1+Mn}\\
    \vdots & \vdots & & \vdots & \vdots\\
    \mu^{(t)}_{k+n-1} & \mu^{(t)}_{k+n-1+M} & \dots & \mu^{(t)}_{k+n-1+M(n-1)} & \mu^{(t)}_{k+n-1+Mn}\\
    1 & z & \dots & z^{n-1} & z^n
  \end{vmatrix}
  ,&
  n=1, 2, 3, \dots,
\end{align*}
where
\begin{equation}\label{eq:det-tau}
  \tau^{(k, t)}_0\coloneq 1,\quad
  \tau^{(k, t)}_n\coloneq |\mu^{(t)}_{k+i+Mj}|_{i, j=0}^{n-1},\quad
  n=1, 2, 3, \dots.
\end{equation}

From the theory of biorthogonal polynomials,
it is shown that one of the pair of the $(M, 1)$-biorthogonal polynomials
$\{\phi^{(k, t)}_n(z)\}_{n=0}^\infty$ satisfies the following relations:
\begin{subequations}\label{eq:m1-spectral}
  \begin{align}
    z\phi^{(k+1, t)}_n(z)&=\phi^{(k, t)}_{n+1}(z)+q^{(k, t)}_n \phi^{(k, t)}_n(z),\label{eq:m1-spectral-q}\\
    \phi^{(k-M, t)}_{n+1}(z)&=\phi^{(k, t)}_{n+1}(z)+e^{(k-M, t)}_n \phi^{(k, t)}_n(z),\label{eq:m1-spectral-e}\\
    (z+s^{(t)})\phi^{(k, t+1)}_n(z)&=\phi^{(k, t)}_{n+1}(z)+\tilde q^{(k, t)}_n \phi^{(k, t)}_n(z),\label{eq:m1-spectral-tq}\\
    (z+s^{(t)})\phi^{(k, t+1)}_n(z)&=z\phi^{(k+1, t)}_{n}(z)+a^{(k, t)}_n \phi^{(k, t)}_n(z),\label{eq:m1-spectral-a}\\
    (z+s^{(t)})\phi^{(k, t+1)}_n(z)&=\phi^{(k-M, t)}_{n+1}(z)+b^{(k-M, t)}_n \phi^{(k, t)}_n(z),
\end{align}
\end{subequations}
where
\begin{subequations}\label{eq:ndhtoda-tau}
  \begin{alignat}{3}
    &q^{(k, t)}_n=\frac{\tau^{(k, t)}_n \tau^{(k+1, t)}_{n+1}}{\tau^{(k, t)}_{n+1}\tau^{(k+1, t)}_n},&\quad
    &e^{(k, t)}_n=\frac{\tau^{(k, t)}_{n+2} \tau^{(k+M, t)}_{n}}{\tau^{(k, t)}_{n+1}\tau^{(k+M, t)}_{n+1}},&\quad
    &\tilde q^{(k, t)}_n=\frac{\tau^{(k, t)}_n \tau^{(k, t+1)}_{n+1}}{\tau^{(k, t)}_{n+1}\tau^{(k, t+1)}_n},\\
    &a^{(k, t)}_n=s^{(t)}\frac{\tau^{(k, t)}_n \tau^{(k+1, t+1)}_{n}}{\tau^{(k+1, t)}_{n}\tau^{(k, t+1)}_n},&\quad
    &b^{(k, t)}_n=\frac{\tau^{(k+M, t)}_n \tau^{(k, t+1)}_{n+1}}{\tau^{(k, t)}_{n+1}\tau^{(k+M, t+1)}_n}.
  \end{alignat}
\end{subequations}
The relations~\eqref{eq:m1-spectral} are called \emph{spectral transformations}.
The compatibility conditions for \eqref{eq:m1-spectral} give the
recurrence relations
\begin{subequations}\label{eq:ndhtoda}
  \begin{alignat}{2}
    &\tilde q^{(k, t)}_n=q^{(k, t)}_n+a^{(k, t)}_n,&\quad
    &\tilde q^{(k+M, t)}_n=b^{(k, t)}_n+e^{(k, t)}_n,\\
    &q^{(k, t+1)}_n=q^{(k, t)}_n\frac{\tilde q^{(k+1, t)}_n}{\tilde q^{(k, t)}_n},&
    &b^{(k, t)}_{n+1}=b^{(k, t)}_n\frac{\tilde q^{(k, t)}_{n+1}}{\tilde q^{(k+M, t)}_n},\\
    &a^{(k, t)}_{n+1}=a^{(k, t)}_n\frac{\tilde q^{(k+1, t)}_n}{\tilde q^{(k, t)}_n},&
    &e^{(k, t+1)}_n=e^{(k, t)}_n\frac{\tilde q^{(k, t)}_{n+1}}{\tilde q^{(k+M, t)}_n}.
  \end{alignat}
  for $n=0, 1, 2, \dots$, with the boundary condition
  \begin{equation}
    a^{(k, t)}_0=s^{(t)},\quad
    b^{(k, t)}_0=q^{(k, t)}_0+s^{(t)}
  \end{equation}
  for all $k, t \in \mathbb Z$.
\end{subequations}
In this paper, we call the system~\eqref{eq:ndhtoda} the ndh-Toda lattice,
which is a reduced system of the nonautonomous discrete two-dimensional
Toda lattice.

\subsection{Finite lattice case}

Hereafter, we consider the case in which the finite lattice boundary condition
\begin{equation}\label{eq:ndhtoda-finite-bc}
  \tau^{(k, t)}_n=0 \quad \text{if $n>N$}
\end{equation}
is imposed, where the lattice size $N$ is a positive integer.
This condition implies
\begin{equation}\label{eq:ndhtoda-finite-bc-e}
  e^{(k, t)}_{N-1}=0
\end{equation}
by \eqref{eq:ndhtoda-tau}.
Hence, from \eqref{eq:m1-spectral-e}, we obtain the relation
\begin{equation}\label{eq:invariant-N}
  \phi^{(k+M, t)}_N(z)=\phi^{(k, t)}_N(z).
\end{equation}
Note that, although we omitted the relations for $\{\psi^{(k, t)}_n(z)\}_{n=0}^{N}$,
there is also the ``dual'' relation
\begin{equation*}
  \psi^{(k+1, t)}_N(z)=\psi^{(k, t+1)}_N(z)=\psi^{(k, t)}_N(z).
\end{equation*}
By using the $N\times N$ bidiagonal matrices
\begin{alignat*}{2}
  &R^{(k, t)}\coloneq
  \begin{pmatrix}
    q^{(k, t)}_0 & 1\\
    & q^{(k, t)}_1 & 1\\
    && \ddots & \ddots\\
    &&& \ddots & 1\\
    &&&& q^{(k, t)}_{N-1}
  \end{pmatrix},&\quad&
  L^{(k, t)}\coloneq
  \begin{pmatrix}
    1\\
    e^{(k, t)}_0 & 1\\
    & e^{(k, t)}_1 & \ddots\\
    && \ddots & \ddots\\
    &&& e^{(k, t)}_{N-2} & 1
  \end{pmatrix},\\
  &\tilde R^{(k, t)}\coloneq
  \begin{pmatrix}
    \tilde q^{(k, t)}_0 & 1\\
    & \tilde q^{(k, t)}_1 & 1\\
    && \ddots & \ddots\\
    &&& \ddots & 1\\
    &&&& \tilde q^{(k, t)}_{N-1}
  \end{pmatrix},
\end{alignat*}
and the $N$-dimensional vectors
\begin{equation*}
  \bm\phi^{(k, t)}(z)\coloneq
  \begin{pmatrix}
    \phi^{(k, t)}_0(z)\\
    \phi^{(k, t)}_1(z)\\
    \vdots\\
    \phi^{(k, t)}_{N-1}(z)
  \end{pmatrix},\quad
  \bm\phi^{(k, t)}_N(z)\coloneq
  \begin{pmatrix}
    0\\
    \vdots\\
    0\\
    \phi^{(k, t)}_{N}(z)
  \end{pmatrix},
\end{equation*}
the spectral transformations~\eqref{eq:m1-spectral-q}--\eqref{eq:m1-spectral-tq} are
written as
\begin{subequations}
  \begin{align}
    z\bm\phi^{(k+1, t)}(z)&=R^{(k, t)}\bm \phi^{(k, t)}(z)+\bm \phi^{(k, t)}_N(z),\label{eq:sp-mat-R}\\
    \bm\phi^{(k-M, t)}(z)&=L^{(k-M, t)}\bm\phi^{(k, t)}(z),\label{eq:sp-mat-L}\\
    (z+s^{(t)})\bm\phi^{(k, t+1)}(z)&=\tilde R^{(k, t)}\bm \phi^{(k, t)}(z)+\bm \phi^{(k, t)}_N(z),
  \end{align}
\end{subequations}
respectively.
We thus have
\begin{subequations}\label{eq:m1-spectral-compat}
  \begin{align}
    z(z+s^{(t)})\bm\phi^{(k+1, t+1)}(z)
    &=R^{(k, t+1)}\tilde R^{(k, t)}\bm\phi^{(k, t)}(z)+R^{(k, t+1)}\bm\phi^{(k, t)}_N(z)+(z+s^{(t)})\bm\phi^{(k, t+1)}_N(z)\nonumber\\
    &=\tilde R^{(k+1, t)}R^{(k, t)}\bm\phi^{(k, t)}(z)+\tilde R^{(k+1, t)}\bm\phi^{(k, t)}_N(z)+z\bm\phi^{(k+1, t)}_N(z),\\
    (z+s^{(t)})\bm\phi^{(k, t+1)}(z)
    &=L^{(k, t+1)}\tilde R^{(k+M, t)}\bm\phi^{(k+M, t)}(z)+L^{(k, t+1)}\bm\phi^{(k+M, t)}_N(z)\nonumber\\
    &=\tilde R^{(k, t)}L^{(k, t)}\bm\phi^{(k+M, t)}(z)+\bm\phi^{(k, t)}_N(z).
  \end{align}
\end{subequations}
We should remark that the relation
\begin{equation*}
  a^{(k, t)}_n
  =a^{(k, t)}_{n-1}\frac{\tilde q^{(k+1, t)}_{n-1}}{\tilde q^{(k, t)}_{n-1}}
  =(\tilde q^{(k, t)}_{n-1}-q^{(k, t)}_{n-1})\frac{\tilde q^{(k+1, t)}_{n-1}}{\tilde q^{(k, t)}_{n-1}}
  =\tilde q^{(k+1, t)}_{n-1}-q^{(k, t+1)}_{n-1}
\end{equation*}
holds by the recurrence relations \eqref{eq:ndhtoda}.
Thus the relation~\eqref{eq:m1-spectral-a} for $n=N$ is rewritten as
\begin{equation*}
  q^{(k, t+1)}_{N-1}\phi^{(k, t)}_{N}(z)+(z+s^{(t)})\phi^{(k, t+1)}_{N}(z)
  =\tilde q^{(k+1, t)}_{N-1}\phi^{(k, t)}_{N}(z)+z\phi^{(k+1, t)}_{N}(z).
\end{equation*}
This relation is equivalent to
\begin{equation*}
  R^{(k, t+1)}\bm\phi^{(k, t)}_N(z)+(z+s^{(t)})\bm\phi^{(k, t+1)}_N(z)
  =\tilde R^{(k+1, t)}\bm\phi^{(k, t)}_N(z)+z\bm\phi^{(k+1, t)}_N(z).
\end{equation*}
Further, $L^{(k, t+1)}\bm\phi^{(k+M, t)}_N(z)=\bm\phi^{(k+M, t)}_N(z)=\bm\phi^{(k, t)}_N(z)$
holds by \eqref{eq:invariant-N}.
Hence, the compatibility conditions for \eqref{eq:m1-spectral-compat}
are simply written as
\begin{equation}\label{eq:ndhtoda-mat}
  R^{(k, t+1)}\tilde R^{(k, t)}=\tilde R^{(k+1, t)}R^{(k, t)},\quad
  L^{(k, t+1)}\tilde R^{(k+M, t)}=\tilde R^{(k, t)}L^{(k, t)}.
\end{equation}

Now consider the upper Hessenberg matrix
\begin{equation*}
  H^{(k, t)}\coloneq L^{(k, t)}R^{(k+M-1, t)}R^{(k+M-2, t)}\dots R^{(k, t)}.
\end{equation*}
By using \eqref{eq:ndhtoda-mat}, we find
\begin{align*}
  H^{(k, t+1)}\tilde R^{(k, t)}
  &=L^{(k, t+1)}R^{(k+M-1, t+1)}\dots R^{(k+2, t+1)}R^{(k+1, t+1)}R^{(k, t+1)}\tilde R^{(k, t)}\\
  &=L^{(k, t+1)}R^{(k+M-1, t+1)}\dots R^{(k+2, t+1)}R^{(k+1, t+1)}\tilde R^{(k+1, t)}R^{(k, t)}\\
  &=L^{(k, t+1)}R^{(k+M-1, t+1)}\dots R^{(k+2, t+1)}\tilde R^{(k+2, t)}R^{(k+1, t)}R^{(k, t)}\\
  &=\dots\\
  &=L^{(k, t+1)}\tilde R^{(k+M, t)}R^{(k+M-1, t)}\dots R^{(k+2, t)}R^{(k+1, t)}R^{(k, t)}\\
  &=\tilde R^{(k, t)}L^{(k, t)}R^{(k+M-1, t)}\dots R^{(k+2, t)}R^{(k+1, t)}R^{(k, t)}\\
  &=\tilde R^{(k, t)}H^{(k, t)}.
\end{align*}
Hence, the ndh-Toda lattice~\eqref{eq:ndhtoda} with
the finite lattice boundary condition~\eqref{eq:ndhtoda-finite-bc}
also can compute iterations of similarity transformations of the upper Hessenberg
matrices as with another nonautonomous version of the discrete hungry Toda lattice
studied in the previous paper~\cite{maeda2017nuh}.

Next, we construct solutions to the finite ndh-Toda lattice~\eqref{eq:ndhtoda} with
\eqref{eq:ndhtoda-finite-bc}.
Suppose that all the eigenvalues $z_0, z_1, \dots, z_{N-1}$ of $H^{(k, t)}$ are simple.
Note that these eigenvalues are directly related to
a polynomial constructed by $\phi^{(k, t)}_N(z), \phi^{(k+1, t)}_N(z), \dots$, and $\phi^{(k+M-1, t)}_N(z)$; see also Appendix~\ref{sec:char-polyn-hk}.
By the result of the previous paper~\cite{maeda2017nuh},
there exist some complex-valued functions $w^{(m)}_0, w^{(m)}_1, \dots, w^{(m)}_{N-1}$
satisfying $w^{(m)}_r=w^{(m \bmod M)}_r$ for $m \in \mathbb Z$, $r=0, 1, \dots, N-1$,
such that the moments of $\mathcal L^{(0, 0)}$ are written as
\begin{equation*}
  \mu^{(0)}_m=\mathcal L^{(0, 0)}[z^m]=\sum_{r=0}^{N-1} w^{(m)}_r z_r^{m/M}.
\end{equation*}
We assume that $z_r \ne 0$ for $r=0, 1, \dots, N-1$.
The relation~\eqref{eq:time-evol-lf} yields
\begin{align}
  \mu^{(t)}_m
  &=\mathcal L^{(0, t)}[z^m]\nonumber\\
  &=\mathcal L^{(0, 0)}\left[z^m \prod_{j=0}^{t-1}(z+s^{(j)})\right]\nonumber\\
  &=\mathcal L^{(0, 0)}\left[z^{m+t}+\sum_{i=1}^{t} \left(\sum_{0\le j_0<j_1<\dots<j_{i-1}\le t-1}\prod_{l=0}^{i-1}s^{(j_l)}\right)z^{m+t-i}\right]\nonumber\\
  &=\sum_{r=0}^{N-1}\tilde w_r^{(m, t)} z_r^{m/M},\label{eq:moment-repr}
\end{align}
where
\begin{gather*}
  \tilde w_r^{(m, t)}\coloneq \sum_{i=0}^{t} w_{r}^{(m+i)}z_r^{i/M}\tilde s_{t-i}^{(t)},\\
  \tilde s_i^{(t)}\coloneq
  \begin{dcases*}
    1& if $i=0$,\\
    \sum_{0\le j_0<j_1<\dots<j_{i-1}\le t-1}\prod_{l=0}^{i-1}s^{(j_l)}& if $i=1, 2, \dots, t$,\\
    0& otherwise.
  \end{dcases*}
\end{gather*}
By definition, $\tilde w_r^{(m, 0)}=w_r^{(m)}$ holds.
It is clear that $\tilde w^{(m, t)}_r=\tilde w^{(m\bmod M, t)}_r$ also holds.

Substituting the moment representation~\eqref{eq:moment-repr} into
the determinant~\eqref{eq:det-tau}, we find
\begin{equation*}
  \tau^{(k, t)}_n=\det(\tilde V_n\mathcal D^{(k)}V^{(k, t)}_n),
\end{equation*}
where
\begin{gather*}
  \tilde V_n\coloneq
  \begin{pmatrix}
    1 & 1 & \dots & 1\\
    z_0 & z_1 & \dots & z_{N-1}\\
    \vdots & \vdots & & \vdots\\
    z_0^{n-1} & z_1^{n-1} & \dots & z_{N-1}^{n-1}
  \end{pmatrix},\\
  \mathcal D^{(k)}\coloneq \diag\left(z_0^{k/M}, z_1^{k/M}, \dots, z_{N-1}^{k/M}\right),\\
  V^{(k, t)}_n\coloneq
  \begin{pmatrix}
    \tilde w^{(k, t)}_0 & \tilde w^{(k+1, t)}_0 z_0^{1/M} & \dots & \tilde w^{(k+n-1, t)}_0 z_0^{(n-1)/M}\\
    \tilde w^{(k, t)}_1 & \tilde w^{(k+1, t)}_1 z_1^{1/M} & \dots & \tilde w^{(k+n-1, t)}_1 z_1^{(n-1)/M}\\
    \vdots & \vdots & & \vdots\\
    \tilde w^{(k, t)}_{N-1} & \tilde w^{(k+1, t)}_{N-1} z_{N-1}^{1/M} & \dots & \tilde w^{(k+n-1, t)}_{N-1} z_{N-1}^{(n-1)/M}
  \end{pmatrix}.
\end{gather*}
Let us introduce the minors of $V^{(k, t)}_{N}$:
\begin{equation*}
  V^{(k, t)}
  \begin{pmatrix}
    r_0, r_1, \dots, r_{n-1}\\
    c_0, c_1, \dots, c_{n-1}
  \end{pmatrix}\coloneq
  \begin{vmatrix}
    \tilde w^{(k+c_0, t)}_{r_0} z_{r_0}^{c_0/M} & \tilde w^{(k+c_1, t)}_{r_0} z_{r_0}^{c_1/M} & \dots & \tilde w^{(k+c_{n-1}, t)}_{r_0} z_{r_0}^{c_{n-1}/M}\\
    \tilde w^{(k+c_0, t)}_{r_1} z_{r_1}^{c_0/M} & \tilde w^{(k+c_1, t)}_{r_1} z_{r_1}^{c_1/M} & \dots & \tilde w^{(k+c_{n-1}, t)}_{r_1} z_{r_1}^{c_{n-1}/M}\\
    \vdots & \vdots & & \vdots\\
    \tilde w^{(k+c_0, t)}_{r_{n-1}} z_{r_{n-1}}^{c_0/M} & \tilde w^{(k+c_1, t)}_{r_{n-1}} z_{r_{n-1}}^{c_1/M} & \dots & \tilde w^{(k+c_{n-1}, t)}_{r_{n-1}} z_{r_{n-1}}^{c_{n-1}/M}
  \end{vmatrix}.
\end{equation*}
We allow $c_0, c_1, \dots, c_{n-1}$ to take values larger than $N-1$.
By definition, we have
\begin{equation}\label{eq:V-t-to-0}
  V^{(k, t)}
  \begin{pmatrix}
    r_0\\
    c_0
  \end{pmatrix}
  =\tilde w^{(k+c_0, t)}_{r_0}z_{r_0}^{c_0/M}
  =\sum_{i=0}^t w_{r_0}^{(k+c_0+i)}z_{r_0}^{(c_0+i)/M}\tilde s^{(t)}_{t-i}
  =\sum_{i=0}^t V^{(k, 0)}
  \begin{pmatrix}
    r_0\\
    c_0+i
  \end{pmatrix}\tilde s^{(t)}_{t-i}
\end{equation}
and
\begin{equation}\label{eq:V-repr}
  V^{(k, t)}
  \begin{pmatrix}
    r_0, r_1, \dots, r_{n-1}\\
    c_0, c_1, \dots, c_{n-1}
  \end{pmatrix}
  =
  \begin{vmatrix}
    V^{(k, t)}
    \begin{pmatrix}
      r_0\\
      c_0
    \end{pmatrix} &
    V^{(k, t)}
    \begin{pmatrix}
      r_0\\
      c_1
    \end{pmatrix} & \dots &
    V^{(k, t)}
    \begin{pmatrix}
      r_0\\
      c_{n-1}
    \end{pmatrix} \\
    V^{(k, t)}
    \begin{pmatrix}
      r_1\\
      c_0
    \end{pmatrix} &
    V^{(k, t)}
    \begin{pmatrix}
      r_1\\
      c_1
    \end{pmatrix} & \dots &
    V^{(k, t)}
    \begin{pmatrix}
      r_1\\
      c_{n-1}
    \end{pmatrix} \\
    \vdots & \vdots && \vdots\\
    V^{(k, t)}
    \begin{pmatrix}
      r_{n-1}\\
      c_0
    \end{pmatrix} &
    V^{(k, t)}
    \begin{pmatrix}
      r_{n-1}\\
      c_1
    \end{pmatrix} & \dots &
    V^{(k, t)}
    \begin{pmatrix}
      r_{n-1}\\
      c_{n-1}
    \end{pmatrix}
  \end{vmatrix}.
\end{equation}

Applying the Binet--Cauchy formula and the expansion formula for
the Vandermonde determinant, we obtain
\begin{equation}\label{eq:ndhtoda-tau-expanded}
  \tau^{(k, t)}_n=\sum_{0\le r_0<r_1<\dots<r_{n-1}\le N-1}V^{(k, t)}
  \begin{pmatrix}
    r_0, r_1, \dots, r_{n-1}\\
    0, 1, \dots, n-1
  \end{pmatrix}
  \prod_{j=0}^{n-1}z_{r_j}^{k/M} \prod_{0\le i<j\le n-1}(z_{r_j}-z_{r_i}).
\end{equation}
Here, from \eqref{eq:V-t-to-0} and \eqref{eq:V-repr}, we have
\begin{align*}
  &\phantom{{}={}}V^{(k, t)}
  \begin{pmatrix}
    r_0, r_1, \dots, r_{n-1}\\
    0, 1, \dots, n-1
  \end{pmatrix}\\
  &=
  \begin{vmatrix}
    \sum_{i=0}^t V^{(k, 0)}
    \begin{pmatrix}
      r_0\\
      i
    \end{pmatrix}\tilde s^{(t)}_{t-i} & \sum_{i=0}^{t} V^{(k, 0)}
    \begin{pmatrix}
      r_0\\
      i+1
    \end{pmatrix}\tilde s^{(t)}_{t-i} & \dots & \sum_{i=0}^{t} V^{(k, 0)}
    \begin{pmatrix}
      r_0\\
      i+n-1
    \end{pmatrix}\tilde s^{(t)}_{t-i}\\
    \sum_{i=0}^t V^{(k, 0)}
    \begin{pmatrix}
      r_1\\
      i
    \end{pmatrix}\tilde s^{(t)}_{t-i} & \sum_{i=0}^{t} V^{(k, 0)}
    \begin{pmatrix}
      r_1\\
      i+1
    \end{pmatrix}\tilde s^{(t)}_{t-i} & \dots & \sum_{i=0}^{t} V^{(k, 0)}
    \begin{pmatrix}
      r_1\\
      i+n-1
    \end{pmatrix}\tilde s^{(t)}_{t-i}\\
    \vdots & \vdots && \vdots\\
    \sum_{i=0}^t V^{(k, 0)}
    \begin{pmatrix}
      r_{n-1}\\
      i
    \end{pmatrix}\tilde s^{(t)}_{t-i} & \sum_{i=0}^{t} V^{(k, 0)}
    \begin{pmatrix}
      r_{n-1}\\
      i+1
    \end{pmatrix}\tilde s^{(t)}_{t-i} & \dots & \sum_{i=0}^{t} V^{(k, 0)}
    \begin{pmatrix}
      r_{n-1}\\
      i+n-1
    \end{pmatrix}\tilde s^{(t)}_{t-i}\\
  \end{vmatrix}\\
  &=\sum_{c_0=0}^t \sum_{c_1=1}^{t+1}\dots \sum_{c_{n-1}=n-1}^{t+n-1}
  \begin{vmatrix}
    V^{(k, 0)}
    \begin{pmatrix}
      r_0\\
      c_0
    \end{pmatrix} & V^{(k, 0)}
    \begin{pmatrix}
      r_0\\
      c_1
    \end{pmatrix} & \dots & V^{(k, 0)}
    \begin{pmatrix}
      r_0\\
      c_{n-1}
    \end{pmatrix}\\
    V^{(k, 0)}
    \begin{pmatrix}
      r_1\\
      c_0
    \end{pmatrix} & V^{(k, 0)}
    \begin{pmatrix}
      r_1\\
      c_1
    \end{pmatrix} & \dots & V^{(k, 0)}
    \begin{pmatrix}
      r_1\\
      c_{n-1}
    \end{pmatrix}\\
    \vdots & \vdots & & \vdots\\
    V^{(k, 0)}
    \begin{pmatrix}
      r_{n-1}\\
      c_0
    \end{pmatrix} & V^{(k, 0)}
    \begin{pmatrix}
      r_{n-1}\\
      c_1
    \end{pmatrix} & \dots & V^{(k, 0)}
    \begin{pmatrix}
      r_{n-1}\\
      c_{n-1}
    \end{pmatrix}\\
  \end{vmatrix}\tilde s^{(t)}_{t-c_0}\tilde s^{(t)}_{t+1-c_1}\dots \tilde s^{(t)}_{t+n-1-c_{n-1}}\\
  &=\sum_{0\le c_0<c_1<\dots<c_{n-1}\le t+n-1} V^{(k, 0)}
  \begin{pmatrix}
    r_0, r_1, \dots, r_{n-1}\\
    c_0, c_1, \dots, c_{n-1}
  \end{pmatrix}\hat s^{(t)}_{(t-c_0, t+1-c_1, \dots, t+n-1-c_{n-1})'},
\end{align*}
where
\begin{gather*}
  \hat s^{(t)}_{(\lambda_0, \lambda_1, \dots, \lambda_{n-1})'}
  \coloneq \sum_{\sigma \in \mathfrak S_n} \sgn \sigma \prod_{j=0}^{n-1} \tilde s^{(t)}_{\lambda_{\sigma(j)}+j-\sigma(j)}
  =
  \begin{vmatrix}
    \tilde s^{(t)}_{\lambda_0} & \tilde s^{(t)}_{\lambda_0+1} & \dots & \tilde s^{(t)}_{\lambda_0+n-1}\\
    \tilde s^{(t)}_{\lambda_1-1} & \tilde s^{(t)}_{\lambda_1} & \dots & \tilde s^{(t)}_{\lambda_1+n-2}\\
    \vdots & \vdots & & \vdots\\
    \tilde s^{(t)}_{\lambda_{n-1}-n+1} & \tilde s^{(t)}_{\lambda_{n-1}-n+2} & \dots & \tilde s^{(t)}_{\lambda_{n-1}}
  \end{vmatrix}.
\end{gather*}
and $\mathfrak S_n$ is the symmetric group on $\{0, 1, \dots, n-1\}$.
Since $\tilde s^{(t)}_0, \tilde s^{(t)}_1, \dots, \tilde s^{(t)}_t$ are the elementary symmetric
polynomials of $s^{(0)}, s^{(1)}, \dots, s^{(t-1)}$, by the Jacobi--Trudi formula,
this is the Schur polynomial for the partition $(\lambda_0, \lambda_1, \dots, \lambda_{n-1})'$, which is the conjugate of the partition $(\lambda_0, \lambda_1, \dots, \lambda_{n-1})$:
\begin{equation*}
  \hat s^{(t)}_{(\lambda_0, \lambda_1, \dots, \lambda_{n-1})'}
  =\sum_{Y}\prod_{j=0}^{t-1} (s^{(j)})^{y_j},
\end{equation*}
where the summation is over all semistandard Young tableaux $Y$ of
the partition $(\lambda_0, \lambda_1, \dots, \lambda_{n-1})'$,
and $y_0, y_1, \dots, y_{n-1}$ are the weights of $Y$.

To derive a sufficient condition for the positivity of $\tau^{(k, t)}_n$,
we discuss recurrence relations among the minors $V^{(k, t)}
\begin{pmatrix}
  r_0, r_1, \dots, r_{n-1}\\
  c_0, c_1, \dots, c_{n-1}
\end{pmatrix}
$.
The followings are readily derived by definition:
\begin{subequations}\label{eq:V-one-two}
\begin{align}
  V^{(k, 0)}
  \begin{pmatrix}
    r_0\\
    c_0
  \end{pmatrix}&=w_{r_0}^{(k+c_0)}z_{r_0}^{c_0/M},\\
  V^{(k, 0)}
  \begin{pmatrix}
    r_0, r_1\\
    c_0, c_1
  \end{pmatrix}
  &=
  \begin{vmatrix}
    V^{(k, 0)}
    \begin{pmatrix}
      r_0\\
      c_0
    \end{pmatrix} &
    V^{(k, 0)}
    \begin{pmatrix}
      r_0\\
      c_1
    \end{pmatrix}\\
    V^{(k, 0)}
    \begin{pmatrix}
      r_1\\
      c_0
    \end{pmatrix} &
    V^{(k, 0)}
    \begin{pmatrix}
      r_1\\
      c_1
    \end{pmatrix}
  \end{vmatrix}.
\end{align}
\end{subequations}
For $n\ge 3$, the Jacobi identity yields
\begin{align}
  \phantom{{}={}}V^{(k, 0)}
  \begin{pmatrix}
    r_0, r_1, \dots, r_{n-3}, r_{n-2}, r_{n-1}\\
    c_0, c_1, \dots, c_{n-3}, c_{n-2}, c_{n-1}
  \end{pmatrix}
  =\frac{\begin{vmatrix}
      V^{(k, 0)}
  \begin{pmatrix}
    r_0, r_1, \dots, r_{n-3}, r_{n-2}\\
    c_0, c_1, \dots, c_{n-3}, c_{n-2}
  \end{pmatrix} &
  V^{(k, 0)}
  \begin{pmatrix}
    r_0, r_1, \dots, r_{n-3}, r_{n-2}\\
    c_0, c_1, \dots, c_{n-3}, c_{n-1}
  \end{pmatrix}\\
  V^{(k, 0)}
  \begin{pmatrix}
    r_0, r_1, \dots, r_{n-3}, r_{n-1}\\
    c_0, c_1, \dots, c_{n-3}, c_{n-2}
  \end{pmatrix} &
  V^{(k, 0)}
  \begin{pmatrix}
    r_0, r_1, \dots, r_{n-3}, r_{n-1}\\
    c_0, c_1, \dots, c_{n-3}, c_{n-1}
  \end{pmatrix}
    \end{vmatrix}}{V^{(k, 0)}
  \begin{pmatrix}
    r_0, r_1, \dots, r_{n-3}\\
    c_0, c_1 \dots, c_{n-3}
  \end{pmatrix}}\label{eq:jacobi}
\end{align}
if $V^{(k+c_0, t)}
\begin{pmatrix}
  r_0, r_1, \dots, r_{n-3}\\
  c_0, c_1, \dots, c_{n-3}
\end{pmatrix}\ne 0$.
Further, by definition,
\begin{equation*}
  V^{(k, 0)}
  \begin{pmatrix}
    r_0, r_1, \dots, r_{n-1}\\
    c_0, c_1, \dots, c_{n-1}
  \end{pmatrix}
  =V^{(k+c_0, 0)}
  \begin{pmatrix}
    r_0, r_1, \dots, r_{n-1}\\
    0, c_1-c_0, \dots, c_{n-1}-c_0
  \end{pmatrix}\prod_{j=0}^{n-1}z_{r_j}^{c_0/M},
\end{equation*}
holds.
Hence, the followings give a sufficient condition for $\tau^{(k, t)}_n>0$
for all $k=0, 1, \dots, M-1$, $t \in \mathbb Z$ and $n=1, 2, \dots, N-1$:
\begin{enumerate}
\item All the parameters $z_r$, $s^{(t)}$, $w^{(k)}_r$ are real and the real $M$th root $z_r^{1/M}$ are chosen;
\item $0<z_0^{1/M}<z_1^{1/M}<\dots<z_{N-1}^{1/M}$;
\item $s^{(t)}\ge 0$ for all $t \in \mathbb Z$;
\item $w^{(k)}_r>0$ for all $k=0, 1, \dots, M-1$ and $r=0, 1, \dots, N-1$;
\item The inequality
  \begin{equation*}
    V^{(k, 0)}
    \begin{pmatrix}
      r_0\\
      0
    \end{pmatrix}
    V^{(k, 0)}
    \begin{pmatrix}
      r_1\\
      c_1
    \end{pmatrix}>
    V^{(k, 0)}
    \begin{pmatrix}
      r_1\\
      0
    \end{pmatrix}
    V^{(k, 0)}
    \begin{pmatrix}
      r_0\\
      c_1
    \end{pmatrix},
  \end{equation*}
  i.e.
  \begin{equation}\label{eq:w-two-condition}
    w^{(k)}_{r_0}w^{(k+c_1)}_{r_1}z_{r_1}^{c_1/M}>w^{(k+c_1)}_{r_0}w^{(k)}_{r_1}z_{r_0}^{c_1/M}
  \end{equation}
  holds for all $k=0, 1, \dots, M-1$, $c_1=1, 2, \dots, M-1$ and pairs of indices $(r_0, r_1)$ satisfying
  $0\le r_0<r_1<N-1$;
\item The inequality
  \begin{multline}\label{eq:v-two-condition}
  V^{(k, 0)}
  \begin{pmatrix}
    r_0, r_1\dots, r_{n-3}, r_{n-2}\\
    0, c_1, \dots, c_{n-3}, c_{n-2}
  \end{pmatrix}
  V^{(k, 0)}
  \begin{pmatrix}
    r_0, r_1\dots, r_{n-3}, r_{n-1}\\
    0, c_1, \dots, c_{n-3}, c_{n-1}
  \end{pmatrix}\\
  >
  V^{(k, 0)}
  \begin{pmatrix}
    r_0, r_1\dots, r_{n-3}, r_{n-1}\\
    0, c_1, \dots, c_{n-3}, c_{n-2}
  \end{pmatrix}
  V^{(k, 0)}
  \begin{pmatrix}
    r_0, r_1\dots, r_{n-3}, r_{n-2}\\
    0, c_1, \dots, c_{n-3}, c_{n-1}
  \end{pmatrix}
  \end{multline}
  holds for all $k=0, 1, \dots, M-1$, $n=3, 4, \dots, N-1$, $n$-tuples
  $(r_0, r_1, \dots, r_{n-1})$ satisfying $0\le r_0<r_1<\dots<r_{n-1}\le N-1$
  and $(n-1)$-tuples $(c_1, c_2, \dots, c_{n-1})$ satisfying
  $0<c_1<c_2<\dots<c_{n-1}$,
  $c_1\le M$ and $c_{i}-c_{i-1}\le M$, $i=2, 3, \dots, n-1$.
\end{enumerate}
We should remark on the condition~\eqref{eq:w-two-condition} that,
since $w^{(k+M)}_r=w^{(k)}_r$ and $z_{r_1}>z_{r_0}$, this condition implies
$w^{(k)}_{r_0}w^{(k+c_1)}_{r_1}z_{r_1}^{c_1/M}>w^{(k+c_1)}_{r_0}w^{(k)}_{r_1}z_{r_0}^{c_1/M}$
for all positive integers $c_1$.

\subsection{Ultradiscretization}
Let us consider the transformations of variables
$q^{(k, t)}_n=\rme^{-Q^{(k, t)}_n/\epsilon}$,
$\tilde q^{(k, t)}_n=\rme^{-\tilde Q^{(k, t)}_n/\epsilon}$,
$e^{(k, t)}_n=\rme^{-E^{(k, t)}_n/\epsilon}$,
$a^{(k, t)}_n=\rme^{-A^{(k, t)}_n/\epsilon}$,
$b^{(k, t)}_n=\rme^{-B^{(k, t)}_n/\epsilon}$
and $s^{(t)}=\rme^{-S^{(t)}/\epsilon}$,
where $\epsilon$ is a positive parameter.
Since there is an ultradiscretization formula
\begin{equation*}
  \lim_{\epsilon\to +0}-\epsilon \log(p_1 \rme^{-C_1/\epsilon}+p_2\rme^{-C_2/\epsilon})=\min(C_1, C_2),
\end{equation*}
where $p_1$ and $p_2$ are positive numbers,
applying these transformations and taking a limit $\epsilon \to +0$ yields
\begin{subequations}\label{eq:nuhtoda}
  \begin{alignat}{2}
    \tilde Q^{(k, t)}_n&=\min(Q^{(k, t)}_n, A^{(k, t)}_n),&\quad
    \tilde Q^{(k+M, t)}_n&=\min(B^{(k, t)}_n, E^{(k, t)}_n),\\
    Q^{(k, t+1)}_n&=Q^{(k, t)}_n-\tilde Q^{(k, t)}_n+\tilde Q^{(k+1, t)}_n,&
    A^{(k, t)}_{n+1}&=A^{(k, t)}_n-\tilde Q^{(k, t)}_n+\tilde Q^{(k+1, t)}_n,\\
    E^{(k, t+1)}_n&=E^{(k, t)}_n-\tilde Q^{(k+M, t)}_n+\tilde Q^{(k, t)}_{n+1},&
    B^{(k, t)}_{n+1}&=B^{(k, t)}_n-\tilde Q^{(k+M, t)}_n+\tilde Q^{(k, t)}_{n+1},
  \end{alignat}
  for $n=0, 1, 2, \dots$ with the boundary condition
  \begin{equation}
    A^{(k, t)}_0=S^{(t)},\quad
    B^{(k, t)}_0=\min(Q^{(k, t)}_0, S^{(t)})
  \end{equation}
  for all $k, t \in \mathbb Z$.
\end{subequations}
In addition, we also impose the finite lattice condition
corresponding to \eqref{eq:ndhtoda-finite-bc-e}:
\begin{equation}\label{eq:nuhtoda-finite-bc-e}
  E^{(k, t)}_{N-1}=+\infty.
\end{equation}
The derived ultradiscrete system~\eqref{eq:nuhtoda} coincides with the time evolution equation
of finite Toda representation~\eqref{eq:evol-eq-ca-toda}.

A solution to the ultradiscrete system~\eqref{eq:nuhtoda} with the finite lattice
condition~\eqref{eq:nuhtoda-finite-bc-e} is constructed from
the solution~\eqref{eq:ndhtoda-tau} and \eqref{eq:ndhtoda-tau-expanded} to
the ndh-Toda lattice~\eqref{eq:ndhtoda}.
Consider the transformations of variables
$z_r=p_r\rme^{-Z_r/\epsilon}$,
$w_r^{(m)}=\rme^{-W_r^{(m)}/\epsilon}$ and
$V^{(k, t)}
\begin{pmatrix}
  r_0, r_1, \dots, r_{n-1}\\
  c_0, c_1, \dots, c_{n-1}
\end{pmatrix}=\exp\left(-\mathcal V^{(k, t)}
\begin{pmatrix}
  r_0, r_1, \dots, r_{n-1}\\
  c_0, c_1, \dots, c_{n-1}
\end{pmatrix}/\epsilon\right)
$ and the limit $\epsilon \to +0$,
where $p_r$ is a positive constant satisfying $p_r<p_{r+1}$ if $Z_r=Z_{r+1}$.
Since we assumed that the inequality $0<z_0<z_1<\dots<z_{N-1}$ holds,
the new variables $Z_r$ must satisfy $Z_0\ge Z_1\ge \dots\ge Z_{N-1}$.
To apply the transformations, $\tau^{(k, t)}_n$ must be positive;
i.e., by \eqref{eq:w-two-condition} and \eqref{eq:v-two-condition},
the followings must be satisfied:
\begin{itemize}
\item The inequality
  \begin{equation}\label{eq:w-condition}
    W_{r_0}^{(k)}+W_{r_1}^{(k+c_1)}+\frac{c_1 Z_{r_1}}{M}\le W_{r_0}^{(k+c_1)}+W_{r_1}^{(k)}+\frac{c_1 Z_{r_0}}{M}
  \end{equation}
  holds for all $k=0, 1, \dots, M-1$, $c_1=1, 2, \dots, M-1$ and
  pairs of indices $(r_0, r_1)$ satisfying $0\le r_0<r_1\le N-1$;
\item The inequality
  \begin{multline}\label{eq:u-v-two-condition}
  \mathcal V^{(k, 0)}
  \begin{pmatrix}
    r_0, r_1, \dots, r_{n-3}, r_{n-2}\\
    0, c_1, \dots, c_{n-3}, c_{n-2}
  \end{pmatrix}+\mathcal V^{(k, 0)}
  \begin{pmatrix}
    r_0, r_1, \dots, r_{n-3}, r_{n-1}\\
    0, c_1, \dots, c_{n-3}, c_{n-1}
  \end{pmatrix}\\
  \le \mathcal V^{(k, 0)}\begin{pmatrix}
    r_0, r_1, \dots, r_{n-3}, r_{n-1}\\
    0, c_1, \dots, c_{n-3}, c_{n-2}
  \end{pmatrix}+\mathcal V^{(k, 0)}
  \begin{pmatrix}
    r_0, r_1, \dots, r_{n-3}, r_{n-2}\\
    0, c_1, \dots, c_{n-3}, c_{n-1}
  \end{pmatrix}
  \end{multline}
  holds for all $k=0, 1, \dots, M-1$, $n=2, 3, \dots, N-1$, $n$-tuples
  $(r_0, r_1, \dots, r_{n-1})$ satisfying $0\le r_0<r_1<\dots<r_{n-1}\le N-1$
  and $(n-1)$-tuples $(c_1, c_2, \dots, c_{n-1})$ satisfying
  $0<c_1<c_2<\dots<c_{n-1}$,
  $c_1\le M$ and $c_{i}-c_{i-1}\le M$, $i=2, 3, \dots, n-1$.
\end{itemize}
We should remark that the following formula holds:
\begin{equation*}
  \lim_{\epsilon \to +0} -\epsilon \log(p_1 \rme^{-C_1/\epsilon}-p_2\rme^{-C_2/\epsilon})=C_1\quad \text{if $A<B$ or $A=B$ and $p_1>p_2>0$}.
\end{equation*}
Hence, under the conditions above, by \eqref{eq:V-one-two}, we obtain
\begin{align*}
  \mathcal V^{(k, 0)}
  \begin{pmatrix}
    r_0\\
    c_0
  \end{pmatrix}&=W_{r_0}^{(k+c_0)}+\frac{c_0 Z_{r_0}}{M},\\
  \mathcal V^{(k, 0)}
  \begin{pmatrix}
    r_0, r_1\\
    c_0, c_1
  \end{pmatrix}
  &=W_{r_0}^{(k+c_0)}+W_{r_1}^{(k+c_1)}+\frac{c_0 Z_{r_0}+c_1 Z_{r_1}}{M}.
\end{align*}
Further, \eqref{eq:jacobi} yields the relation
\begin{align*}
  \mathcal V^{(k, 0)}
  \begin{pmatrix}
    r_0, r_1, \dots, r_{n-3}, r_{n-2}, r_{n-1}\\
    c_0, c_1, \dots, c_{n-3}, c_{n-2}, c_{n-1}
  \end{pmatrix}
  &=\mathcal V^{(k, 0)}
  \begin{pmatrix}
    r_0, r_1, \dots, r_{n-3}, r_{n-2}\\
    c_0, c_1, \dots, c_{n-3}, c_{n-2}
  \end{pmatrix}+
  V^{(k, 0)}
  \begin{pmatrix}
    r_0, r_1, \dots, r_{n-3}, r_{n-1}\\
    c_0, c_1, \dots, c_{n-3}, c_{n-1}
  \end{pmatrix}\nonumber\\
  &\qquad
  -\mathcal V^{(k, 0)}
  \begin{pmatrix}
    r_0, r_1, \dots, r_{n-3}\\
    c_0, c_1, \dots, c_{n-3}
  \end{pmatrix}
\end{align*}
for $n \ge 3$.
Hence, we obtain
\begin{align*}
  \mathcal V^{(k, 0)}
  \begin{pmatrix}
    r_0, r_1, r_2\\
    c_0, c_1, c_2
  \end{pmatrix}
  &=\mathcal V^{(k, 0)}
  \begin{pmatrix}
    r_0, r_1\\
    c_0, c_1
  \end{pmatrix}+
  \mathcal V^{(k, 0)}
  \begin{pmatrix}
    r_0, r_2\\
    c_0, c_2
  \end{pmatrix}-
  \mathcal V^{(k, 0)}
  \begin{pmatrix}
    r_0\\
    c_0
  \end{pmatrix}\\
  &=\sum_{j=0}^2 \left(W_{r_j}^{(k+c_j)}+\frac{c_j Z_{r_j}}{M}\right).
\end{align*}
In general, the following equation is proved by induction on $n$:
\begin{equation*}
  \mathcal V^{(k, 0)}
  \begin{pmatrix}
    r_0, r_1, \dots, r_{n-1}\\
    c_0, c_1, \dots, c_{n-1}
  \end{pmatrix}=\sum_{j=0}^{n-1}\left(W_{r_j}^{(k+c_j)}+\frac{c_j Z_{r_j}}{M}\right).
\end{equation*}
Substituting this result into the inequality~\eqref{eq:u-v-two-condition},
we obtain the simpler condition
\begin{equation*}
  W^{(k+c_0)}_{r_0}+W^{(k+c_1)}_{r_1}+\frac{c_0 Z_{r_0}+c_1Z_{r_1}}{M}
  \le W^{(k+c_1)}_{r_0}+W^{(k+c_0)}_{r_1}+\frac{c_1 Z_{r_0}+c_0Z_{r_1}}{M}
\end{equation*}
for all pairs $(r_0, r_1)$ satisfying $0\le r_0<r_1\le N-1$, and
all pairs $(c_0, c_1)$ satisfying $0\le c_0<c_1$ and $c_1-c_0\le M$.
This result means that the conditions~\eqref{eq:w-condition}
ensure other conditions~\eqref{eq:u-v-two-condition} hold.

Suppose that all the conditions above are satisfied.
Then, $\tau^{(k, t)}_n$~\eqref{eq:ndhtoda-tau-expanded} is ultradiscretized as
\begin{gather*}
  T^{(k, t)}_n=\min_{\substack{0\le r_0<r_1<\dots<r_{n-1}\le N-1\\ 0\le c_0<c_1<\dots<c_{n-1}\le t+n-1}}\left(\sum_{j=0}^{n-1} \left(W_{r_j}^{(k+c_j)}+\frac{k+Mj+c_j}{M}Z_{r_j}\right)+\hat S^{(t)}_{(t-c_0, t+1-c_1, \dots, t+n-1-c_{n-1})'}\right),\\
  \hat S^{(t)}_{(\lambda_0, \lambda_1, \dots, \lambda_{n-1})'}\coloneq\min_Y \left(\sum_{j=0}^{t-1} y_j S^{(j)}\right),
\end{gather*}
where $\min_Y$ indicates the minimum value over all semistandard Young tableaux $Y$ of
the partition $(\lambda_0, \lambda_1, \dots, \lambda_{n-1})'$,
and $y_0, y_1, \dots, y_{n-1}$ are the weights of $Y$.
Finally, ultradiscretization of \eqref{eq:ndhtoda-tau} yields
\begin{align*}
  Q^{(k, t)}_n&=T^{(k, t)}_n-T^{(k, t)}_{n+1}+T^{(k+1, t)}_{n+1}-T^{(k+1, t)}_n,\\
  E^{(k, t)}_n&=T^{(k, t)}_{n+2}-T^{(k, t)}_{n+1}+T^{(k+M, t)}_{n}-T^{(k+M, t)}_{n+1},\\
  \tilde Q^{(k, t)}_n&=T^{(k, t)}_n-T^{(k, t)}_{n+1}+T^{(k, t+1)}_{n+1}-T^{(k, t+1)}_n,\\
  A^{(k, t)}_n&=T^{(k, t)}_n-T^{(k+1, t)}_{n}+T^{(k+1, t+1)}_{n}-T^{(k, t+1)}_n+S^{(t)},\\
  B^{(k, t)}_n&=T^{(k+M, t)}_n-T^{(k, t)}_{n+1}+T^{(k, t+1)}_{n+1}-T^{(k+M, t+1)}_n.
\end{align*}
For example, the solution corresponding to Fig.~\ref{fig:ex-toda} is given by setting
$Z_0=8$, $Z_1=4$, $Z_2=2$,
$W_0^{(0)}=1$, $W_0^{(1)}=4/3$, $W_0^{(2)}=2/3$,
$W_1^{(0)}=9$, $W_1^{(1)}=26/3$, $W_1^{(2)}=25/3$,
$W_2^{(0)}=15$, $W_2^{(1)}=46/3$, $W_2^{(2)}=47/3$,
and $S^{(t)}=2$ for all $t$.

Finally, let us consider asymptotics for the autonomous case.
If $S^{(t)}=S$ for all $t$, then
\begin{equation*}
  \hat S^{(t)}_{(\lambda_0, \lambda_1, \dots, \lambda_{n-1})'}=\left(\sum_{j=0}^{n-1}\lambda_j\right)S.
\end{equation*}
Hence, in this autonomous case, a solution is given by
\begin{align*}
  T^{(k, t)}_n
  &=\min_{\substack{0\le r_0<r_1<\dots<r_{n-1}\le N-1\\ 0\le c_0<c_1<\dots<c_{n-1}\le t+n-1}}\left(\sum_{j=0}^{n-1} \left(W_{r_j}^{(k+c_j)}+\frac{k+Mj+c_j}{M}Z_{r_j}+(t+j-c_j)S\right)\right)\\
  &=\min_{\substack{0\le r_0<r_1<\dots<r_{n-1}\le N-1\\ 0\le c_0<c_1<\dots<c_{n-1}\le t+n-1}}\left(\sum_{j=0}^{n-1} \left(W_{r_j}^{(k+c_j)}+\frac{k+Mj}{M}Z_{r_j}+(t+j)S+\frac{c_j(Z_{r_j}-MS)}{M}\right)\right).
\end{align*}
Suppose that $Z_{0}\ge \dots \ge Z_{m-1}>MS\ge Z_{m}\ge \dots \ge Z_{N-1}$ holds.
Then, since $W^{(k+M)}_{r_j}=W^{(k)}_{r_j}$,
\begin{align*}
  T^{(k, t+M)}_n-T^{(k, t)}_n&=
  \begin{dcases*}
    \sum_{j=0}^{n-1} Z_{N-n+j} & if $n\le N-m$,\\
    (n-m)MS+\sum_{j=0}^{N-m-1} Z_{m+j} & if $n>N-m$,
  \end{dcases*}
\end{align*}
hold for $t \gg 1$. Hence, we obtain
\begin{equation*}
  Q^{(k, t+M)}_n-Q^{(k, t)}_n=0 \quad \text{for $t \gg 1$},
\end{equation*}
and
\begin{equation*}
  \sum_{k=1}^M Q^{(k, t+M)}_n-\sum_{k=1}^M Q^{(k, t)}_n=0 \quad \text{for $t \gg 1$}.
\end{equation*}
Therefore, in the autonomous case, the size of each soliton converges
to some constant, and the arrangement of positive integers of each soliton changes
in period $M$.
In addition, we have
\begin{equation*}
  T^{(k+M, t)}_n-T^{(k, t)}_n=\sum_{j=0}^{n-1}Z_{N-n+j}
\end{equation*}
for $n\le N-m$ and $t \gg 1$. Hence,
\begin{equation*}
  \sum_{k=1}^M Q^{(k, t)}_n=T^{(1, t)}_n-T^{(1, t)}_{n+1}+T^{(M+1, t)}_{n+1}-T^{(M+1, t)}_n
  =Z_{N-n-1}
\end{equation*}
holds for $n<N-m$ and $t\gg 1$.
Further, we also have
\begin{equation*}
  \sum_{j=0}^{M-1}\tilde Q^{(k, t+j)}_n
  =T^{(k, t)}_n-T^{(k, t)}_{n+1}+T^{(k, t+M)}_{n+1}-T^{(k, t+M)}_n=
  \begin{dcases*}
    Z_{N-n-1} & if $n\le N-m$,\\
    MS & if $n>N-m$,
  \end{dcases*}
\end{equation*}
for $t\gg 1$.
This means that the speed, which is the moving distance from time $t$ to $t+M$,
of the $n$th soliton at sufficiently large time $t$
is given by $\min(Z_{N-n-1}, MS)$.

\section{Concluding remarks}\label{sec:concluding-remarks}

In this paper, we have proposed a novel soliton cellular automaton,
derived two types of time evolution equations to it,
and given particular solutions to the evolution equations.
We have focused on only the time evolution equations and their solutions.
Several important properties of the proposed cellular automaton, e.g. conserved quantities,
linearization~\cite{takagi2005ism,kakei2018lbb},
and relations to the solvable lattice models or Yang--Baxter relation~\cite{inoue2012isb}
should be investigated in detail.
Applications of the ndh-Toda lattice studied in this paper
to numerical algorithms like the one proposed by Fukuda \etal.~\cite{fukuda2012eam}
are also important future works.

In the discussion for the derivation of particular solutions
to both the Euler representation and the finite Toda representation,
variables taking negative or complex values cause a difficulty
for ultradiscretization. The methods used in the previous studies
and this paper are based on analysis to show that dominant terms are positive
under some restricted conditions for parameters.
However, this analysis is complicated and a little hard to perform
in general. For this problem, several methods, i.e. ultradiscretization
for variables taking negative or complex values,
have been proposed~\cite{yajima2006mpa,kasman2006wnn,mimura2009sct}.
Another method is to use permanent solutions
instead of determinant solutions~\cite{takahashi2007uss,nagai2018ups}.
It may be interesting to consider applying these methods for
ultradiscretization of solutions to the systems appearing in this paper.

\section*{Funding}
This work was supported by JSPS KAKENHI Grant Number JP17H02858.

\appendix

\section{Characteristic polynomial of $H^{(k, t)}$}\label{sec:char-polyn-hk}

In this appendix,
we investigate a relationship between $\phi^{(k, t)}_N(z)$ and $H^{(k, t)}$.
By using \eqref{eq:sp-mat-R} and \eqref{eq:sp-mat-L} repeatedly,
we obtain
\begin{align}
  z^M\bm\phi^{(k, t)}(z)
  &=H^{(k, t)}\bm\phi^{(k, t)}(z)\nonumber\\
  &\qquad+\sum_{j=0}^{M-1} L^{(k, t)}R^{(k+M-1, t)}R^{(k+M-2, t)}\dots R^{(k+j+1, t)}z^j\bm\phi^{(k+j, t)}_N(z).\label{eq:eig-H-sep}
\end{align}
Substituting $z=z\rme^{-2\pi\rmi \nu/M}$, $\nu=0, 1, \dots, M-1$, into \eqref{eq:eig-H-sep},
we obtain
\begin{align}
  &\phantom{{}={}}z^M\bm\phi^{(k, t)}(z\rme^{-2\pi\rmi \nu/M})\nonumber\\
  &=H^{(k, t)}\bm\phi^{(k, t)}(z\rme^{-2\pi\rmi \nu/M})\nonumber\\
  &\qquad+\sum_{j=0}^{M-1} L^{(k, t)}R^{(k+M-1, t)}R^{(k+M-2, t)}\dots R^{(k+j+1, t)}z^j\rme^{-2\pi\rmi \nu j/M}\bm\phi^{(k+j, t)}_N(z\rme^{-2\pi\rmi \nu/M}).\label{eq:eig-H-sep-e}
\end{align}
Consider the linear combination of \eqref{eq:eig-H-sep-e}:
\begin{align}\label{eq:eig-H-sep-e-l}
  &\phantom{{}={}}z^M\sum_{\nu=0}^{M-1}\hat w^{(k, t)}_{\nu}\bm\phi^{(k, t)}(z\rme^{-2\pi\rmi \nu/M})\nonumber\\
  &=H^{(k, t)}\sum_{\nu=0}^{M-1}\hat w^{(k, t)}_{\nu}\bm\phi^{(k, t)}(z\rme^{-2\pi\rmi \nu/M})\nonumber\\
  &\qquad+\sum_{j=0}^{M-1} L^{(k, t)}R^{(k+M-1, t)}R^{(k+M-2, t)}\dots R^{(k+j+1, t)}z^j\sum_{\nu=0}^{M-1}\hat w^{(k, t)}_{\nu}\rme^{-2\pi\rmi \nu j/M}\bm\phi^{(k+j, t)}_N(z\rme^{-2\pi\rmi \nu/M}),
\end{align}
where $\hat w^{(k, t)}_0, \hat w^{(k, t)}_1, \dots, \hat w^{(k, t)}_{M-1} \in \mathbb C$ are constants.
If there exist a value $z_r \in \mathbb C$ and
a nonzero vector $(\hat w^{(k, t)}_{r, 0}, \hat w^{(k, t)}_{r, 1}, \dots, \hat w^{(k, t)}_{r, M-1})$
satisfying
\begin{equation}\label{eq:phi-le-z}
  \sum_{\nu=0}^{M-1}\hat w^{(k, t)}_{r, \nu}\rme^{-2\pi\rmi \nu j/M}\phi^{(k+j, t)}_N(z_r\rme^{-2\pi\rmi \nu/M})=0,\quad j=0, 1, \dots, M-1,
\end{equation}
then equation \eqref{eq:eig-H-sep-e-l} gives
\begin{equation*}
  z_r^M\sum_{\nu=0}^{M-1}\hat w^{(k, t)}_{r, \nu}\bm\phi^{(k, t)}(z_r\rme^{-2\pi\rmi \nu/M})
  =H^{(k, t)}\sum_{\nu=0}^{M-1}\hat w^{(k, t)}_{r, \nu}\bm\phi^{(k, t)}(z_r\rme^{-2\pi\rmi \nu/M}).
\end{equation*}
This equation means that $z_r^M$ and $\sum_{\nu=0}^{M-1}\hat w^{(k, t)}_{r, \nu}\bm\phi^{(k, t)}(z_r\rme^{-2\pi\rmi \nu/M})$ are an eigenvalue and a corresponding eigenvector
of the matrix $H^{(k, t)}$, respectively.
To exist nonzero solutions for the linear system \eqref{eq:phi-le-z},
the value $z_r$ must be a zero of the following polynomial of degree $MN$:
\begin{align*}
  &\Phi^{(k, t)}_N(z)\coloneq\\
  &\begin{vmatrix}
    \phi^{(k, t)}_N(z) & \phi^{(k, t)}_N(z\rme^{-2\pi\rmi/M}) & \dots & \phi^{(k, t)}_N(z\rme^{-2\pi\rmi(M-1)/M})\\
    \phi^{(k+1, t)}_N(z) & \rme^{-2\pi\rmi/M}\phi^{(k+1, t)}_N(z\rme^{-2\pi\rmi/M}) & \dots & \rme^{-2\pi\rmi(M-1)/M}\phi^{(k+1, t)}_N(z\rme^{-2\pi\rmi(M-1)/M})\\
    \vdots & \vdots &  & \vdots\\
    \phi^{(k+M-1, t)}_N(z) & \rme^{-2\pi\rmi(M-1)/M}\phi^{(k+M-1, t)}_N(z\rme^{-2\pi\rmi/M}) & \dots & \rme^{-2\pi\rmi(M-1)^2/M}\phi^{(k+M-1, t)}_N(z\rme^{-2\pi\rmi(M-1)/M})
  \end{vmatrix}.
\end{align*}
Note that, since the determinant is a multilinear alternating map,
it is readily shown that
\begin{equation*}
  \Phi^{(k, t)}_N(z\rme^{-2\pi \rmi/M})
  =(-1)^{M-1}\rme^{2\pi\rmi/M}\rme^{2\pi\rmi\cdot 2/M}\dots \rme^{2\pi\rmi(M-1)/M}\Phi^{(k, t)}_N(z)
  =\Phi^{(k, t)}_N(z).
\end{equation*}
This implies that the relation
\begin{equation*}
  \Phi^{(k, t)}_N(z\rme^{-2\pi \rmi \nu/M})=\Phi^{(k, t)}_N(z),\quad
  \nu=0, 1, \dots, M-1,
\end{equation*}
holds.
Hence, $\Phi^{(k, t)}_N(z^{1/M})$ is a polynomial of degree $N$
and its zeros are the eigenvalues of $H^{(k, t)}$.

In addition to the $(M, 1)$-orthogonality relation~\eqref{eq:m1-biorth},
let us consider the \emph{discrete} $(M, 1)$-orthogonality relation
\begin{equation}\label{eq:m1-orthogonality-N}
  \mathcal L^{(k, t)}[z^{m}\psi^{(k, t)}_N(z^M)]=0,\quad
  m=0, 1, 2, \dots.
\end{equation}
Suppose that all the zeros of $\Phi^{(k, t)}_N(z)$ and $\psi^{(k, t)}_N(z)$
are simple.
Then, as discussed in the previous paper~\cite{maeda2017nuh},
$\psi^{(k, t)}_N(z)$ is the characteristic polynomial of $H^{(k, t)}$.
Therefore, it is shown that the relation
\begin{equation*}
  C_{M, n}^{-1}\Phi^{(k, t)}_N(z)=\psi^{(k, t)}_N(z^M)
\end{equation*}
holds, where the constant $C_{M, n}$ is the leading coefficient of $\Phi^{(k, t)}_N(z)$,
which is calculated by using the result on the eigenvalues of the discrete Fourier
transform matrix~\cite{mcclellan1972eed}:
\begin{equation*}
  C_{M, n}=(-1)^{(M-1)n+\lfloor(M+2)/4\rfloor}\rmi^{\lfloor (M-1)/4\rfloor}(-\rmi)^{\lfloor(M+1)/4\rfloor}\left(\sqrt{M}\right)^M.
\end{equation*}
The discrete $(M, 1)$-orthogonality relation~\eqref{eq:m1-orthogonality-N} is now
equivalent to
\begin{equation*}
  \mathcal L^{(k, t)}[z^{m}\Phi^{(k, t)}_N(z)]=0,\quad
  m=0, 1, 2, \dots.
\end{equation*}


\end{document}